\documentclass[%
 reprint,
 amsmath,amssymb,
 aps,
twocolumn,
prb,
floatfix,
]{revtex4-2}

\usepackage{graphicx}
\usepackage{dcolumn}
\usepackage{bm}
\usepackage{chemformula} 
\usepackage[T1]{fontenc} 
\usepackage[version=3]{mhchem}
\usepackage{stfloats} 
\usepackage{graphicx} 
\usepackage{placeins}
\usepackage{lastpage}
\usepackage[format=plain,justification=justified,singlelinecheck=false,font={stretch=1.125,small,sf},labelfont=bf,labelsep=space]{caption}
\usepackage{float}
\usepackage{fancyhdr}
\usepackage{tabularx} 
\usepackage{threeparttable}
\usepackage{fnpos}
\usepackage[english]{babel}
\usepackage{array}
\usepackage{droidsans}
\usepackage{charter}
\usepackage[T1]{fontenc}
\usepackage[usenames,dvipsnames]{xcolor}
\usepackage{setspace}
\usepackage[compact]{titlesec}
\usepackage{subcaption}
\usepackage{booktabs}
\definecolor{cream}{RGB}{222,217,201}
\usepackage{xcolor}
\usepackage{braket}

\begin{document}

\preprint{APS/123-QED}

\title{Symmetry-Match Fraction for Active-Space Selection: A Simple Route to Accurate Reaction Energies in VQE/UCCSD}

\author{Maitreyee Sarkar}
 \affiliation{Quantum Information and Computation Lab, Department of Chemistry, Indian Institute of Technology Jodhpur, Rajasthan, India, 342030}
\author{Lisa Roy}%
\affiliation{Quantum Information and Computation Lab, Department of Chemistry, Indian Institute of Technology Jodhpur, Rajasthan, India, 342030}
\author{Akash Popat Gutal}%
\affiliation{Chemical Dynamics Research Group, Department of Chemistry, Indian Institute of Technology Jodhpur, Rajasthan, India, 342030}
\author{Atul Kumar}%
 \email{atulk@iitj.ac.in}
\affiliation{Quantum Information and Computation Lab, Department of Chemistry, Indian Institute of Technology Jodhpur, Rajasthan, India, 342030}
\author{Manikandan Paranjothy}%
 \email{pmanikandan@iitj.ac.in}
\affiliation{Chemical Dynamics Research Group, Department of Chemistry, Indian Institute of Technology Jodhpur, Rajasthan, India, 342030}



\begin{abstract}
Quantum computing is viewed as a promising technology because of its potential for polynomial growth in complexity, in contrast to the exponential growth observed in its classical counterparts. In the current Noisy Intermediate-Scale Quantum (NISQ) era, the Variational Quantum Eigensolver (VQE), a hybrid variational algorithm, is utilized to simulate molecules using qubits and calculate molecular properties. However, simulating a chemical reaction to compute the reaction energy using VQE algorithm has not yet reached chemical accuracy relative to the benchmark computational chemistry methods due to limitations such as the number of qubits, circuit depth, and noise introduced within the model. To address this issue, we propose the definition of different active spaces for studying chemical reactions, incorporating irreducible representations of both the ground and excited states of the molecules by defining the maximum contribution of the excitation terms of the ansatz. Our results demonstrate that this approach achieves chemical accuracy in predicting the reaction energy for various reactions. For all reactions studied, the difference in reaction energies between conventional computational chemistry methods and the quantum-classical hybrid VQE algorithm is less than 1 kcal/mol. Furthermore, our analysis simplifies the process of selecting active spaces and electrons for each reaction, reducing it to a single optimal combination that ensures the chemical accuracy for each reaction.
\end{abstract}

\maketitle


\section{Introduction}
Quantum mechanics underpins a paradigm shift in computing, redefining the landscape of traditional computing industries. In recent years, quantum computing has transpired as a potential technology characterized by polynomial growth in complexity, in contrast to classical computation, where complexity increases exponentially \cite{rieffel2000introduction}. For example, quantum algorithms like Shor’s algorithm \cite{shor1999polynomial} and Grover’s \cite{grover1996fast} algorithm offer monumental advantages in speed and space over their classical counterparts.  These advantages led to applications of quantum algorithms across diverse academic domains within the realms of quantum computing, such as quantum cryptography \cite{pirandola2020advances, bennett1992quantum, bennett1992experimental,nielsen2001quantum}, quantum key distribution \cite{shor2000simple, renner2008security, wolf2021quantum, nielsen2001quantum}, quantum machine learning \cite{schuld2015introduction, ciliberto2018quantum, huang2021power, sharma2024quantum}, quantum finance \cite{orus2019quantum, chang2023prospects}, and quantum chemistry \cite{cao2019quantum, mcardle2020quantum, lanyon2010towards, claudino2022basics, kais2014introduction, fano2019quantum, kassal2011simulating, dirac1929quantum}. Although there are many technological challenges considering the decoherence \cite{brandt1999qubit, barnes1999decoherence, divincenzo1998decoherence, sabale2024facets} and scalability of quantum computation, the academic community has witnessed enormous progress on theoretical as well as experimental fronts. With continued advancements in cutting-edge technology, it is possible to envision a quantum computer solving a variety of complex problems efficiently—problems that would be practically intractable with classical computers.\par


 In the past two decades, one of the promising field for the applications of quantum computation is computational chemistry.  To predict the properties of a molecule,  wave function is required and the difficulties in obtaining the wave function increase exponentially with increasing particle numbers \cite{aspuru2005simulated, armaos2020computational}. Among modern electronic structure theory methods, the coupled cluster singles, doubles, and perturbative triples, i.e., CCSD(T) method, is expressed to be the gold standard because of its higher accuracy in predicting energies. Compared with the experimental and full Configuration Interaction (CI) data, CCSD(T) produces energy values with errors less than a few kJ/mol. However, the scaling of the CCSD(T) method with respect to the size of the system $(N)$ follows $O(N^7)$, limiting its applicability only to a few atom systems. With the advent of quantum computation, quantum computers are becoming essential for understanding molecular properties, as they reduce the problem's complexity from exponential to the polynomial term. \cite{feynman2018simulating}. The unitary coupled cluster (UCC) method is an idea derived from coupled cluster (CC) method of quantum chemistry, as the unitary operations can be naturally executed on a quantum computer. \par

The theoretical foundation for applying quantum computing to computational chemistry began gaining momentum in the late 1990s \cite{zalka1998efficient}. Zalka demonstrated an efficient simulation of the wave function of a many-body system on a quantum computer, with a computational cost comparable to that of conventional classical system simulations \cite{ortiz2001quantum}. One of the best algorithms that can be used for analyzing properties of many-body systems is the quantum phase estimation algorithm (QPEA) \cite{motta2022emerging}. In comparison to conventional computational techniques available for quantum chemistry calculations, the QPEA shows promises of exponential speed-up. However, the main drawback of QPEA is that the circuit depth is very high (i.e., a vast number of quantum gates are required to represent the wavefunction), which is not conceivable without an extensive fault-tolerant quantum computer \cite{parker2020quantum}. \par

The quantum computers available today (also called, Near-term quantum devices) are Noisy Intermediate Scale Quantum Computers (NISQ) \cite{preskill2018quantum} and currently  have an insubstantial number of qubits. This motivated the academic community to work with an amalgamated quantum-classical algorithm, i.e., Variational Quantum Eigensolver (VQE)  \cite{ kandala2017hardware} for computational chemistry calculations. VQE algorithm has gained significant popularity in near-term quantum devices among the various quantum algorithms used for benchmarking and applications in computational chemistry, due to its relatively lower circuit depth compared to fault-tolerant quantum computers \cite{preskill2018quantum, li2019variational}. In order to compute the ground state energy for any molecule, VQE algorithm utilizes variational principle along with a classical gradient-based optimizer\cite{peruzzo2014variational}- a quantum-classical hybrid algorithm. For efficient calculations, a resource conserving approach is to perform the VQE calculations by restricting the Hilbert space, which is defined as active orbitals (which in quantum chemistry is known as active space) \cite{yalouz2021state}. However, due to the limitations in qubit count, circuit depth, and noise in the VQE algorithm, simulating molecular properties does not reach the desired chemical accuracy when compared to traditional computational chemistry methods for various properties of a reaction \cite{liepuoniute2024simulation}. In this article, we efficiently address the question of obtaining the chemical accuracy for a set of reactions using the VQE algorithm and group theory compared to conventional computational chemistry methods.  \par
While the foundational algorithms and quantum chemistry formulations have been extensively studied, there remains a significant gap in translating these techniques into chemically accurate simulations of full reaction processes. In this work, we aim to bridge this gap by focusing not on isolated molecular ground states, but on the accurate computation of reaction energies - a central quantity used in computational chemistry calculations. In recent years, a considerable progress has been made in optimizing variational quantum eigensolver (VQE) circuits using methods such as symmetry exploitation, qubit tapering, and ansatz design. For example, qubit tapering strategies using $Z_{2}$ symmetries \cite{setia2020reducing} and operator compression techniques \cite{cao2022progress} have helped scale simulations to larger systems, while hardware implementations \cite{nam2020ground} have demonstrated the feasibility of running quantum chemistry algorithms on near-term devices. These efforts have laid a critical foundation for resource-efficient quantum simulations. 
However, much of this prior work has focused on individual molecules or hardware-centric performance improvements, with relatively less emphasis on the end-to-end simulation of chemical reactions and the accurate prediction of reaction energetics. Addressing this, we introduce a symmetry-informed framework that integrates irreducible representations (irreps) of molecular point groups into excitation filtering. Rather than using symmetry solely for circuit optimization, we employ it as a fundamental criterion to select which active spaces to include in the quantum simulation. This approach significantly reduces the active space search space while maintaining chemical relevance and accuracy. \par

In the present work, we develop a group-theoretic protocol that filters excitations based on the symmetry properties of the involved electronic states to assign probabilistic weights. Thereafter, we apply this framework to five chemically and industrially relevant reactions- including halogenation, Sabatier methanation, and ammonia synthesis- and demonstrate that it achieves reaction energy predictions within 1 kcal/mol, using compact and symmetry-consistent active spaces. More specifically, we employ group-theoretic techniques to assess the relative importance of excitation configurations by analyzing their contributions to the ground-state energy (GSE). This analysis allows us to assign probabilistic weights to excitations based on their symmetry characteristics across all species involved in the reaction. By identifying the symmetry matched fractions (SMFs) across the active spaces of both reactants and products, we simulated five different chemically and industrially significant reactions, as shown below
\begin{itemize}
    \item \ce{H2 + F2} $\rightarrow$  \ce{2HF}
    \item \ce{H2 + Cl2} $\rightarrow $ \ce{2HCl}
    \item \ce{2HI + Cl2} $\rightarrow$  \ce{2HCl + I2}
    \item \ce{3H2 + CO \rightarrow CH4 + H2O}
    \item \ce{3H2 + N2} $\rightarrow$ \ce{2NH3}
\end{itemize}

\section{Methods} \label{method} 
\subsection{VQE Algorithm}

The fundamental concept behind VQE algorithm is based on the variational principle \cite{mcquarrie1997physical}, which determines an upper bound to the ground state energy of a system by using a trial wave function that satisfies the problem's boundary conditions. 
In VQE algorithms, the wave function is represented by a parameterized quantum circuit (PQC)  \cite{ostaszewski2021structure, benedetti2019parameterized, whitfield2011simulation, lee2018generalized}, with the aim of variationally optimizing \cite{romero2018strategies} these parameters to minimize the expectation value of the Hamiltonian. The VQE algorithm consists of several components, each involving key decisions that affect the overall design and cost of the algorithm.\par

\begin{figure}[h]
\centering
\includegraphics[width=6cm]{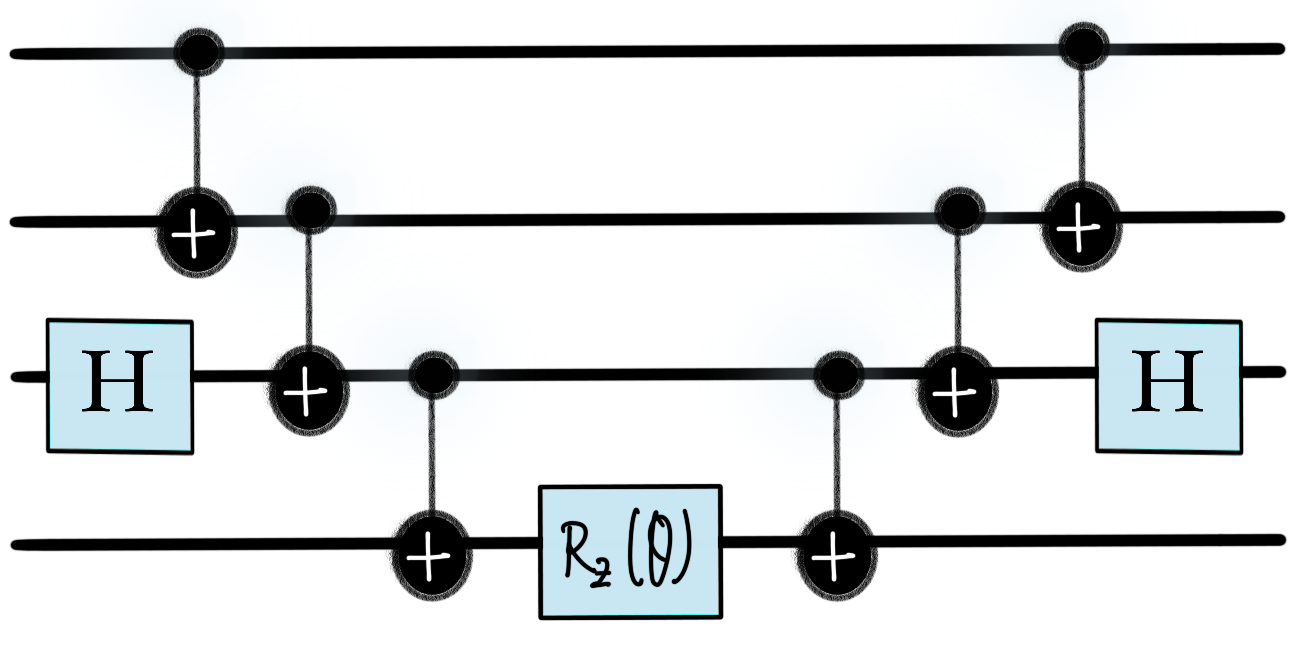}
\caption{Example representation of a random single term ($e^{i\frac{\theta}{2}(Z_1Z_2X_3Z_4)}$) of UCCSD ansatz in as a circuit fragment}
\end{figure}


\subsubsection{Hamiltonian}
 The Hamiltonian operator \cite{alteg2022study} determines the ground state energy of a molecular system. Within the VQE framework, the electronic Hamiltonian is typically expressed in its second-quantized form, as follows

\begin{eqnarray}
\hat{H}_e &=& \sum_{p,q} h_{pq} \hat{a}^{\dag}_p \hat{a}_q + \frac{1}{2} \sum_{p,q,r,s} h_{pqrs} \hat{a}^{\dag}_p \hat{a}^{\dag}_q \hat{a}_r \hat{a}_s
\end{eqnarray}

Here $h_{pq}$ denotes the one-electron integrals, accounting for kinetic energy and nuclear attraction terms, while $h_{pqrs}$ corresponds to the two-electron integrals, capturing electron-electron repulsion. The operators $\hat{a}^{\dag}_i$, and $\hat{a}_j$ are the fermionic creation and annihilation operators, respectively, acting on spin-orbital occupation vectors.\par

To enable implementation on a quantum computer, this fermionic Hamiltonian is then mapped to qubit operators using transformation techniques such as Jordan-Wigner, Bravyi-Kitaev, or Parity mapping. These methods translate the second quantized Hamiltonian into a linear combination of tensor products of Pauli operators, which can be directly executed on a quantum hardware. The transformed qubit Hamiltonian can be represented as \par

\begin{eqnarray}
\hat{H}_e &=& \sum_\alpha h_\alpha P_\alpha; P \in \{I,X,Y,Z\}
\end{eqnarray}

\begin{figure}[h]
\centering

\begin{subfigure}{0.3\textwidth}
    \centering
    \includegraphics[width=\linewidth,height=0.8\textheight,keepaspectratio]{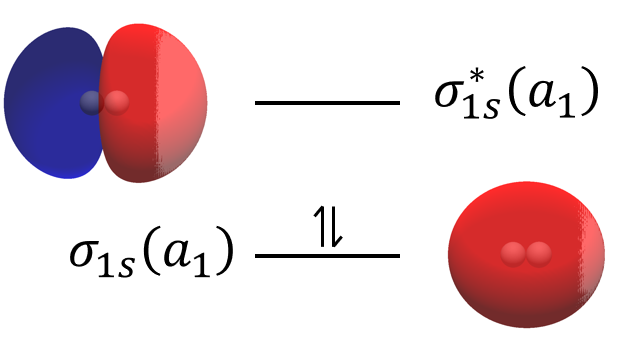}
    \caption{Molecular orbitals}
\end{subfigure}
\begin{subfigure}{0.46\textwidth}
    \centering
    \includegraphics[width=\linewidth,height=0.8\textheight,keepaspectratio]{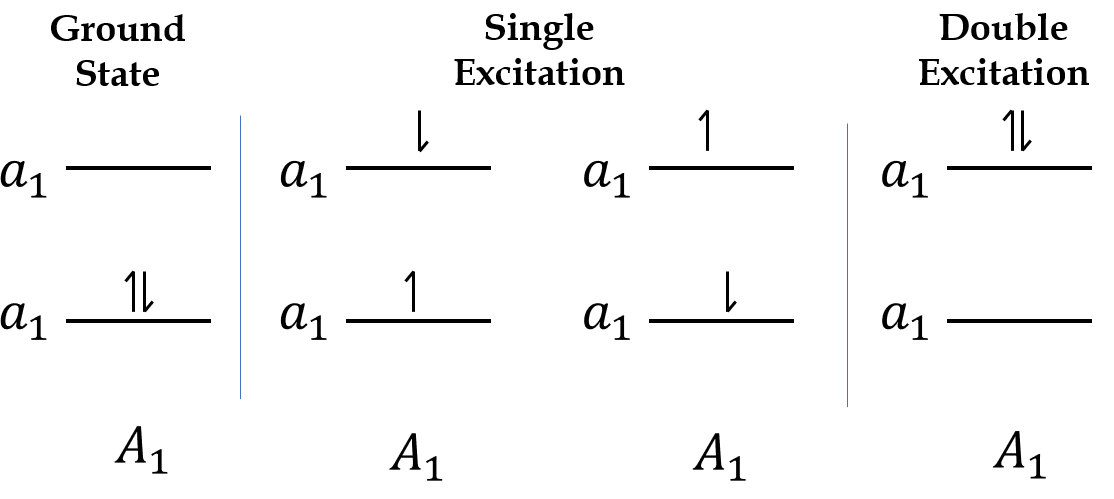}
    \caption{Irreducible representations}
\end{subfigure}
\caption{Different descriptions of \ce{H$_2$} molecule.}
\label{fig:main}
\end{figure}

\begin{figure*}[h]
    \centering
    \begin{subfigure}{0.8\textwidth}
        \centering
        \includegraphics[width=0.6\textwidth]{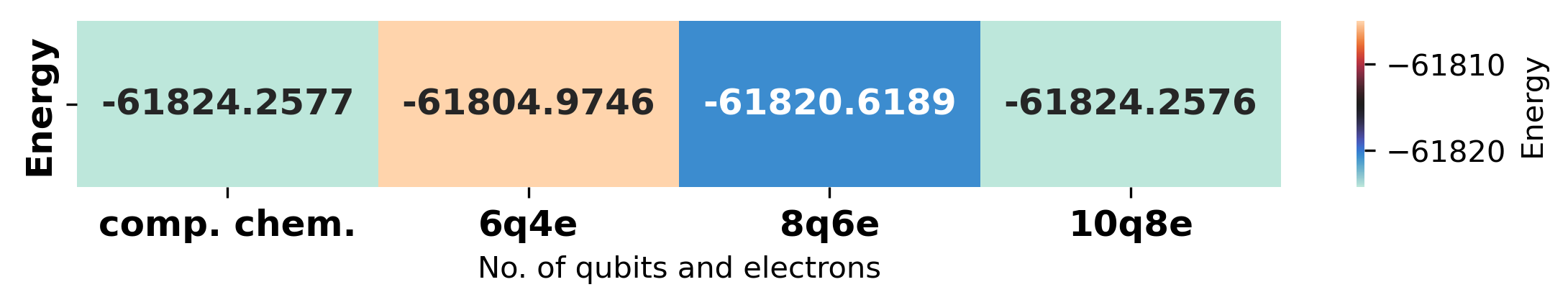}
        \caption{Energy values for HF Molecule}
        \label{fig:hf_energy}
    \end{subfigure}
    
    
    \begin{subfigure}{0.8\textwidth}
        \centering
        \includegraphics[width=0.6\textwidth]{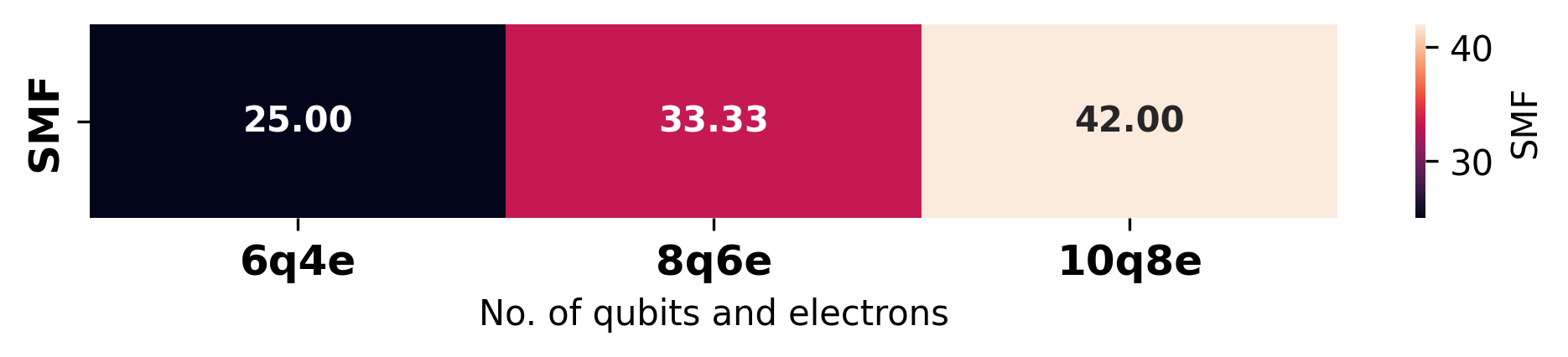}
        \caption{SMF values for HF Molecule}
        \label{fig:hf_prob}
    \end{subfigure}
    
    
    \begin{subfigure}{0.48\textwidth}
        \centering
        \includegraphics[width=0.5\textwidth]{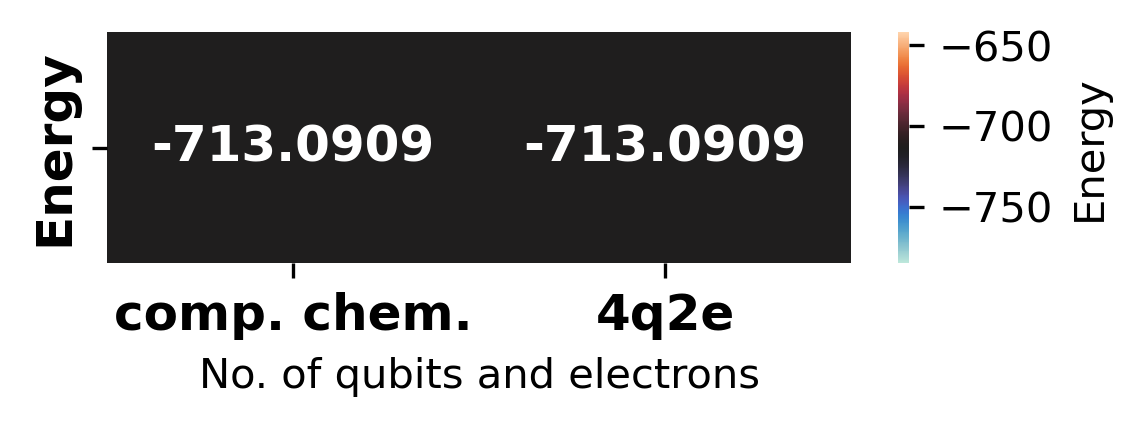}
        \caption{Energy values for \ce{H2} molecule}
        \label{fig:h2_energy}
    \end{subfigure}
    \begin{subfigure}{0.48\textwidth}
        \centering
        \includegraphics[width=0.45\textwidth]{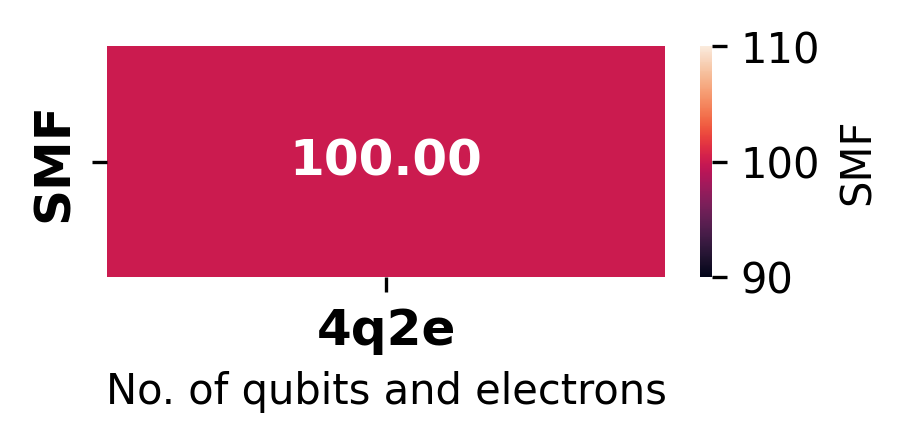}
        \caption{SMF values for \ce{H2} molecule}
        \label{fig:h2_prob}
    \end{subfigure}
    
    
    \begin{subfigure}{0.8\textwidth}
        \centering
        \includegraphics[width=0.9\textwidth]{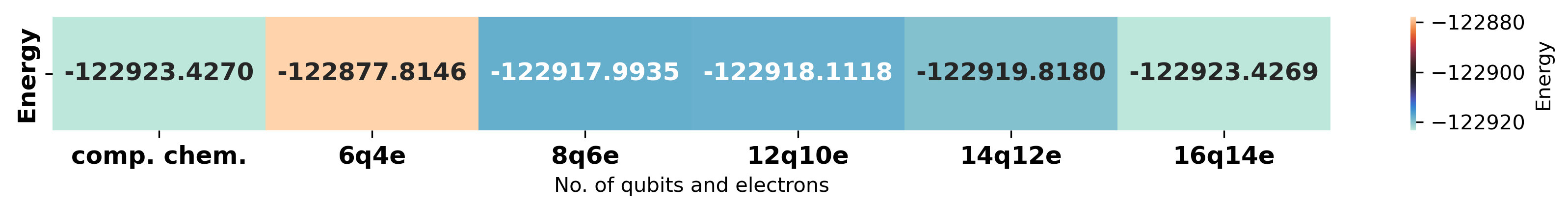}
        \caption{Energy values for \ce{F2} molecule}
        \label{fig:f2_energy}
    \end{subfigure}
    
    
    \begin{subfigure}{0.8\textwidth}
        \centering
        \includegraphics[width=0.8\textwidth]{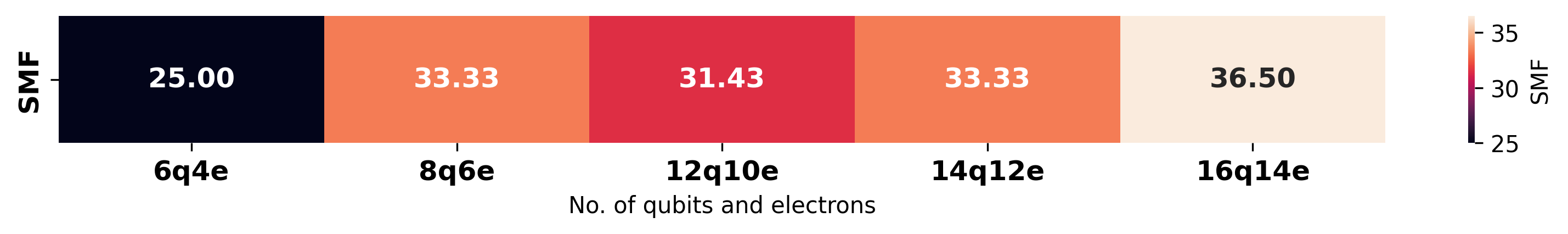}
        \caption{SMF values for \ce{F2} molecule}
        \label{fig:f2_prob}
    \end{subfigure}
    
    
    \begin{subfigure}{0.8\textwidth}
        \centering
        \includegraphics[width=0.9\textwidth]{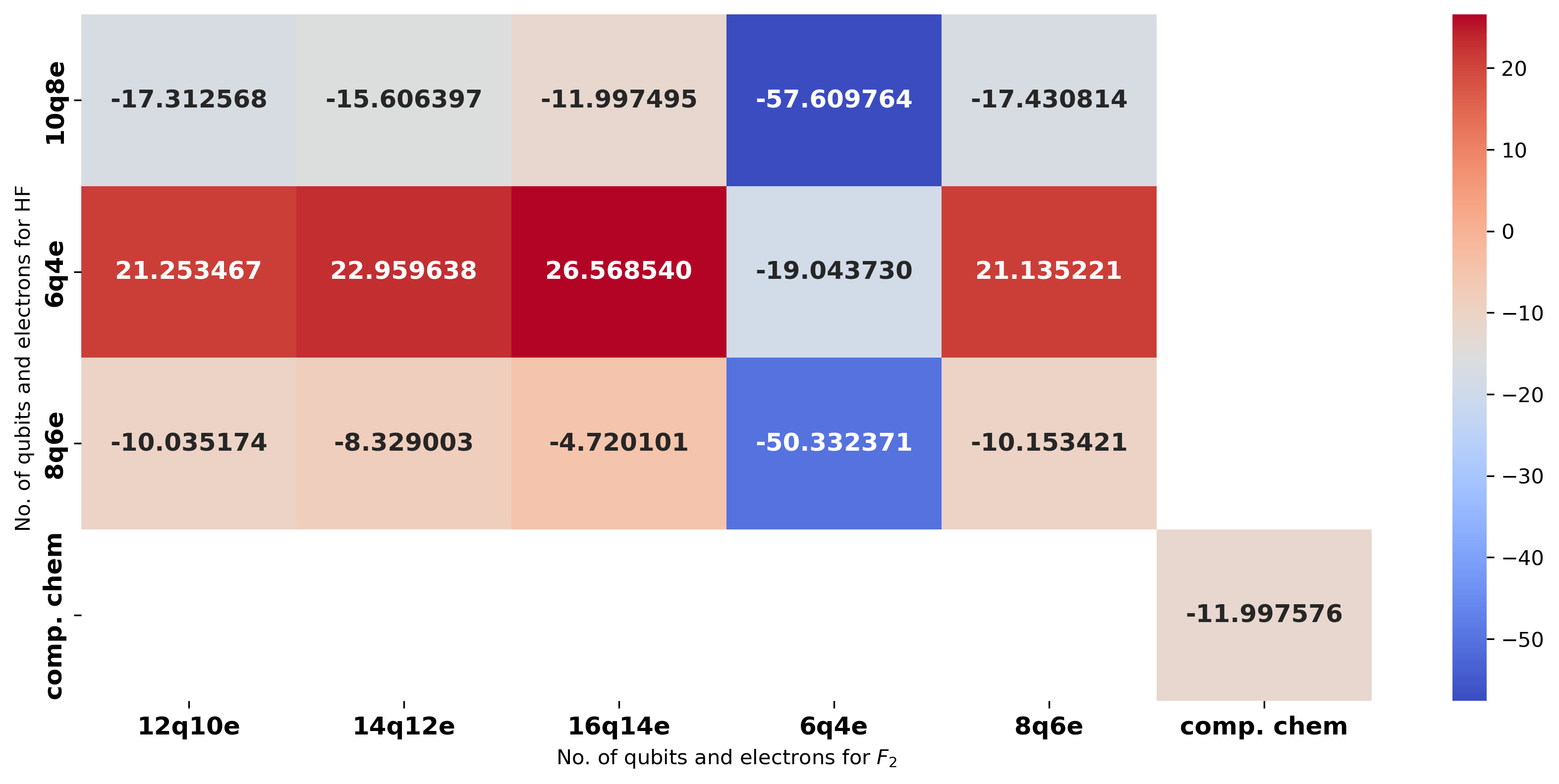}
        \caption{Combinations of reaction energies for the generation of HF from \ce{H2} and \ce{F2}}
        \label{fig:reaction1}
    \end{subfigure}
    
    \caption{Energy and SMF values for reactants, products and reaction energies  of Reaction 1.}
    \label{fig:main}
\end{figure*}

\subsubsection{Ansatz}

The Unitary Coupled Cluster (UCC) ansatz \cite{grimsley2019trotterized} is one of the most widely adopted approaches for determining the ground state energy within the VQE framework. The general form of the UCC ansatz is given by

\begin{eqnarray}
    \ket{\psi} = e^{T - T^\dag} \ket{\psi_0}
\end{eqnarray}

where, $T$ denotes the excitation operator, and $\ket{\psi_0}$  is a reference state, typically the Hartree-Fock state. To make the ansatz variational, $T$ is expressed as a linear combination of excitation operators weighted by trainable parameters $\theta_{i}$ such that 

\begin{subequations}
    \begin{eqnarray}
        T &=& \sum_i \theta_i T_i, \label{subeq:1}
    \end{eqnarray}
    \begin{eqnarray}
        T^{\dag} &=& \sum_i \theta_i T^{\dag}_i, \label{subeq:2}
    \end{eqnarray}
\end{subequations}
Since the exponential of the full operator $T$-$T^{\dag}$ is difficult to implement exactly on a quantum computer, we apply second-order Trotterization \cite{romero2018strategies} to decompose $U$ as
\begin{eqnarray}
    U &=& e^{T-T^\dag} = (\prod_i e^{\frac{\theta_i}{2}(T_i-T^\dag_i)})^2
\end{eqnarray}
The individual excitation operators $T_{i}$ are constructed using fermionic creation and annihilation operators, and are subsequently transformed into qubit operators using mappings such as Jordan-Wigner, Bravyi-Kitaev, or Parity mapping, as done for the Hamiltonian. This results in a unitary evolution operator expressed as a product of exponentials of Pauli strings
\begin{eqnarray}
    U &=& (\prod_i \prod_j e^{i{\frac{\theta_i}{2}(P_{ij})}})^2
\end{eqnarray}
where $P_{ij}$ are the mapped Pauli strings corresponding to each excitation term. 
To illustrate how these exponentials of Pauli strings translate to quantum circuits, a representative operator term such as $e^{i\frac{\theta}{2}(Z_1Z_2X_3Z_4)}$ is shown in Figure 1.

\subsubsection{Classical optimization}
The third and final component of the VQE algorithm is classical optimization, which is essential to this hybrid quantum-classical approach. Classical optimization algorithms are used to iteratively adjust the parameters of the ansatz until a convergence with expected ground state energy is obtained. For the algorithm to be practically viable, the optimizer must be capable of learning to approximate the solution effectively within a tractable number of steps \cite{singh2023benchmarking}. \par

\subsection{Group Theory}
Group theory \cite{cotton1991chemical, mirman1995group, butler2012point} provides a powerful framework for analyzing the symmetry properties of molecules and understanding their influence on molecular behavior. By considering these symmetries, it becomes possible to reduce the dimensionality and complexity of the qubit Hamiltonian used in quantum simulations \cite{setia2020reducing}.
Group-theoretic methods have found diverse applications in both quantum chemistry and quantum computing. For example, Cao et al. \cite{cao2022progress} introduced the sym-VQE method, which utilizes symmetry considerations to reduce the number of operators in the original UCCSD ansatz. Their results demonstrate that despite the reduction, the computed ground state energies remain nearly identical to those obtained using the full operator set. In the UCCSD framework, each excitation term introduces a new variational parameter to the VQE optimization. By restricting excitations to only those that share the same symmetry as the ground state, sym-VQE effectively reduces the number of parameters, and thus the number of operators, without compromising accuracy. This highlights how group theory can serve not only as a tool for understanding molecular structure, but also as a means for enhancing computational efficiency in quantum algorithms.

\section{Determination of Reaction Energies} \label{results} 

\subsection{Computational Details}
\subsubsection{Computational Chemistry Methods}

In this work, we utilize NWChem \cite{valiev2010nwchem} and  Orca \cite{neese2020orca} for the CCSD calculations to optimize and then determine the ground state energies of all the molecules using the frozen-core approximation and STO-3G basis set. Additionally, molecular orbitals were computed for all the molecules with the STO-3G basis set and Avogadro \cite{hanwell2012avogadro} software was used to visualize the molecular orbitals.\par

\subsubsection{Quantum Computation Methods}
To compute the reaction energy using a quantum computer, we utilize Qiskit 0.46.0 \cite{qiskit2024}, Qiskit Nature 0.6.0 \cite{qiskit_nature_2023}, and PySCF 2.0.1 \cite{sun2020recent}. The ground state energies are calculated using the STO-3G basis set, UCCSD ansatz, SLSQP optimizer and the ideal state-vector simulator without a noise model. All results here were obtained with ideal statevector simulations in minimal basis sets. This choice isolates the effect of symmetry-based filtering on variational accuracy without interference from device noise or optimizer stochasticity. Future work will therefore combine the present metric with noise-mitigation strategies and adaptive ansätze on real devices.

\subsection{Mathematical Details of Group Theory}

In our approach, we did not explicitly apply reduction techniques such as qubit tapering or ansatz parameter pruning. Instead, our focus was to investigate whether chemical accuracy in reaction energy calculations could still be achieved when using different active spaces for the reactants and products involved in various chemical reactions. These variations naturally lead to multiple combinations of reaction energies, not all of which fall within the threshold of chemical accuracy. Building on the insight from Cao et al. \cite{cao2022progress}, we note that the dominant contributions to the ground state energy (GSE) typically arise from excitation operators that share the same symmetry as the ground state. Inspired by this, and given that our simulations involve different combinations of qubits (spin orbitals) and electrons across various active spaces, we devised a probabilistic metric to quantify the significance of symmetry-consistent excitations. Specifically, for each active space, we computed the proportion of excitations that match the ground-state symmetry relative to the total number of excitations. This value was used to define a symmetry-matched fraction (SMF) measure reflecting the quality of that active space in capturing the correct electronic structure. For each reactant and product, we then selected the active space configuration with the highest such SMF, and used it in our VQE simulations. This strategy enabled us to achieve chemical accuracy in the computed reaction energies, with errors consistently within the range of $10^{-3}$ kcal/mol. \par

\begin{equation}
    SMF = \frac{N_{\text{excitations with same symmetry as the ground state}}}{N_{\text{excitations}}} \times 100
\end{equation}

By point-group symmetry, any excitation operator whose overall irreducible representation does not contain the totally symmetric irrep yields vanishing matrix elements with the reference determinant. Consequently,  such symmetry-forbidden excitations have identically zero amplitudes in symmetry-preserving perturbative and variational many-body methods. We therefore define the symmetry-match probability $p(A)$ for an active space $A$ as the fraction of singles and doubles within $A$ whose overall irrep contains the totally symmetric irrep. Removing excitations which are symmetry forbidden is cost-neutral in the ideal symmetry-preserving theory and reduces operator/matrix complexity without loss of variational freedom. The symmetry-match fraction provides a simple way to estimate how many excitation operators in a chosen active space are consistent with the overall molecular symmetry. Only these symmetry-allowed excitations can mix with the reference determinant and contribute to the recovery of  correlation energy beyond the mean-field level. Active spaces with a higher symmetry-matched fraction, therefore contain more physically relevant excitation pathways and offer greater variational freedom for UCCSD or VQE optimization. In contrast, spaces with many symmetry-forbidden excitations contain operators that are inactive and cannot lower the total energy. For low-symmetry molecules such as those in $C_1$, all excitations are symmetry-allowed and the symmetry-match fraction correctly returns $p = 1$; in such cases the metric becomes neutral rather than misleading, and active-space ranking proceeds through the size and structure of the excitation manifold. \par
Previous strategies for reducing quantum-chemistry ansatz complexity leverage symmetry and adaptivity in different ways. Qubit tapering exploits discrete symmetries of the fermionic Hamiltonian to eliminate qubits or compress Pauli operators, yielding rigorous resource reductions \cite{bravyi2017tapering}. The ADAPT-VQE framework constructs compact, system-specific circuits by iteratively selecting operators with the largest energy gradients \cite{grimsley2019adaptive}, and later work has emphasized symmetry-adapted operator pools to ensure convergence within the correct irreducible representation \cite{tang2021qubit}. In parallel, natural-orbital and entanglement-based active-space selection methods, such as ASS1ST \cite{khedkar2020extending} or occupation-based orbital pruning \cite{stein2016automated}, identify strongly correlated orbitals via correlated precomputations like CASSCF, NEVPT, or DMRG. Our proposed symmetry-match fraction addresses a different layer of the workflow: it counts only those excitations provably allowed by point-group symmetry, providing a cost-free scalar that ranks candidate active spaces before variational optimization. In this sense, it complements tapering, adaptive ansätze, and entanglement-based orbital selection by filtering out symmetry-forbidden directions at the outset.

\begin{table}
\caption{\label{tab:product_table} Product table\cite{webqc_c2v} for $C_{2v}$ point group}
\begin{tabular}{c|cccc}
\toprule
 $C_{2v}$ & $A_1$ & $A_2$ & $B_1$ & $B_2$\\
\midrule
$A_1$ & $A_1$ & $A_2$ & $B_1$ & $B_2$ \\
$A_2$ & $A_2$ & $A_1$ & $B_2$ & $B_1$ \\
$B_1$ & $B_1$ & $B_2$ & $A_1$ & $A_2$ \\
$B_2$ & $B_2$ & $B_1$ & $A_2$ & $A_1$ \\
\bottomrule
\end{tabular}
\end{table}

 We analyze the direct products of the irreducible representations (irreps) using the product table for the $C_{2v}$ irreps, which is presented in Table I. For instance, Figure 2(a) illustrates that two $1S$ atomic orbitals (AOs) from each hydrogen atom combine to form molecular orbitals (MOs) for the hydrogen molecule, both having the irrep $a_1$, the method for this calculation can be found in Cao et al. \cite{cao2022progress}. All irreducible representations refer to molecular orbitals (MOs) obtained from symmetry-adapted Restricted Hartree-Fock (RHF) calculations. For a linear molecule we follow the standard practice of using the Abelian subgroup to label the MOs. This avoids the complications of non-Abelian irreps while leaving all symmetry assignments and excitation-selection rules unchanged. Figure 2(b) illustrates the irreps of the ground and excited states for a molecule, using the hydrogen molecule as an example. Since the \ce{H2} molecule belongs to the $D_{\infty h}$ point group, we select the abelian subgroup $C_{2v}$ (shown in Table I) for simplicity in calculations.  \par

As a result, the ground state of the \ce{H2} molecule corresponds to the 
irrep $a_1 \times a_1$ = $A_1$. Furthermore, as shown in Figure 2(b), for the \ce{H2} molecule, two single excitations and one double excitation are possible, and all excitation terms possess the $A_1$ irrep (see Table I). Consequently, all excitation terms share the same irrep as the ground state of the \ce{H2} molecule. \par

\subsection{Results}

\subsubsection{\textnormal{\ce{H2 + F2} $\rightarrow$  \ce{2HF}}}

We now proceed to perform a comprehensive analysis of a set of closed-shell chemical reactions to estimate reaction energies using computational chemistry and the VQE algorithm. To begin with, we consider fluorination of hydrogen which is an exothermic reaction. We evaluate the single-point energy for the previously optimized geometries of all the reactants and products using NWChem. As illustrated in Figures 3(a-f), we employ various combinations of spin-orbitals (qubits) and electrons for the VQE algorithm calculations. We use the Jordan–Wigner mapping without any qubit tapering or symmetry-based qubit reduction. Under this mapping, each spatial molecular orbital is represented by two spin orbitals, and each spin orbital corresponds to a single qubit. Therefore, an active space containing nnn spatial orbitals is encoded using 2n2n2n qubits. Since no qubits were removed through tapering, the “number of qubits” is in one-to-one correspondence with the “number of spin orbitals” and thus indirectly with the “number of molecular orbitals.”

We have used heatmap plot for both energy and SMF values for easier visualization of the results. For the VQE energy values plots 3(a), (c), and (e), the first energy value is for the computational chemistry method, and the next ones are for different active spaces, and from the legend, it can be visualized how the energy values are changing with respect to different active spaces. For the SMF value plots 3(b), (d), and (f), by referring to the legends, one can easily visualize that the SMF with the lightest colour has the maximum value. The reaction energy obtained using the CCSD method is used as the reference to estimate the accuracy of the VQE algorithm calculations. Our analysis using VQE algorithm demonstrates that based on spin orbitals and number of electrons, there are a total of 15 combinations required to be optimally analyze and evaluate the reaction energy. Figure 3(g) shows that not all of the 15 possible combinations achieve chemical accuracy. \par

To visualize this result, let us first briefly analyze the active space for the HF molecule. For example, Figure 4 shows three different active spaces (6-qubits or 3 spatial orbitals, 4-electrons; 8-qubits or 4 spatial orbitals, 6-electrons; 10-qubits or 5 spatial orbitals, 8-electrons) for the HF molecule. The active space includes a maximum of five MOs and eight electrons. The energy of the sixth MO (-25.907079 $E_h$) is much lower than that of the fifth and fourth MOs (-1.450018 $E_h$ and -0.561731 $E_h$, respectively), as seen in Figure 4. Therefore, the sixth MO has not been considered for the VQE algorithm calculations. The first active space includes six qubits or 3 spatial orbitals and four electrons due to the two degenerate MOs in the HOMO, and hence the first active space will include three MOs instead of two, as represented in Figures 5 and 4(a). Based on the molecular orbitals, the minimum number of qubits required for \ce{H2}, \ce{F2}, and HF are $4$, $6$, and $6$, respectively. The orbital energies for the \ce{F2} molecule are provided in the supporting information. We further analyze the irreducible representations for all combinations of MOs and electrons for reactants and products for each single and double excitation term to determine how many transitions have the same irreducible representation as the ground state, as mathematically defined in Cao et al. \cite{cao2022progress}. \par

\begin{figure}[h]
\includegraphics[width=7cm, height=7cm]{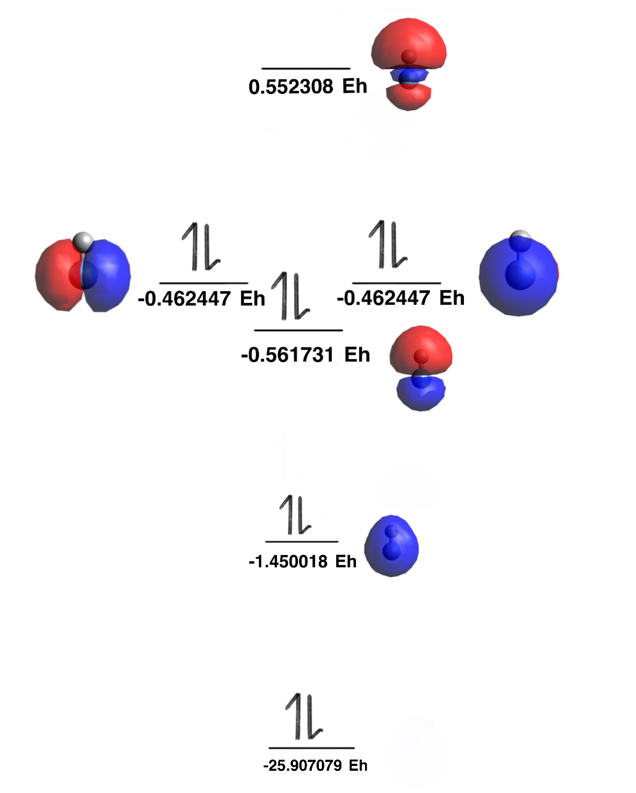}
\caption{MO of HF molecule (image is not upto scale)}
\end{figure}

\begin{figure*}[h]
    \centering
    \begin{subfigure}{0.75\textwidth}
        \centering
        \includegraphics[width=0.7\textwidth]{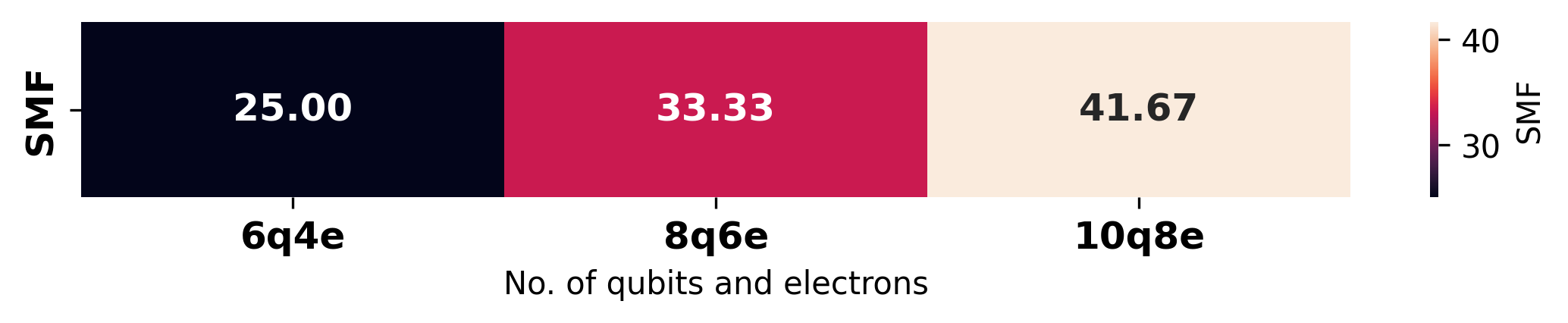}
        \caption{SMF values for HCl Molecule}
        \label{fig:hf_prob}
    \end{subfigure}
    \begin{subfigure}{0.8\textwidth}
        \centering
        \includegraphics[width=0.8\textwidth]{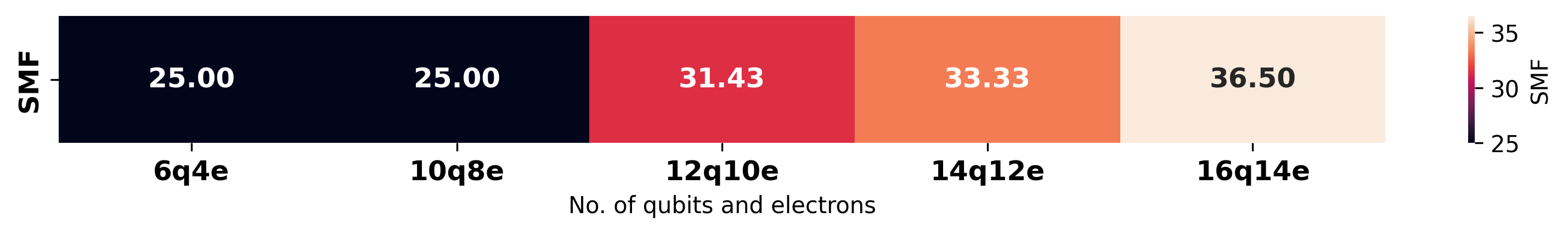}
        \caption{SMF values for \ce{Cl2} molecule}
        \label{fig:f2_prob}
    \end{subfigure}
    \caption{SMF values for reactants and products of Reaction 2.}
    \label{fig:main}
\end{figure*}

\begin{figure*}[h]
    \centering
    \begin{subfigure}{0.8\textwidth}
        \centering
        \includegraphics[width=0.7\textwidth]{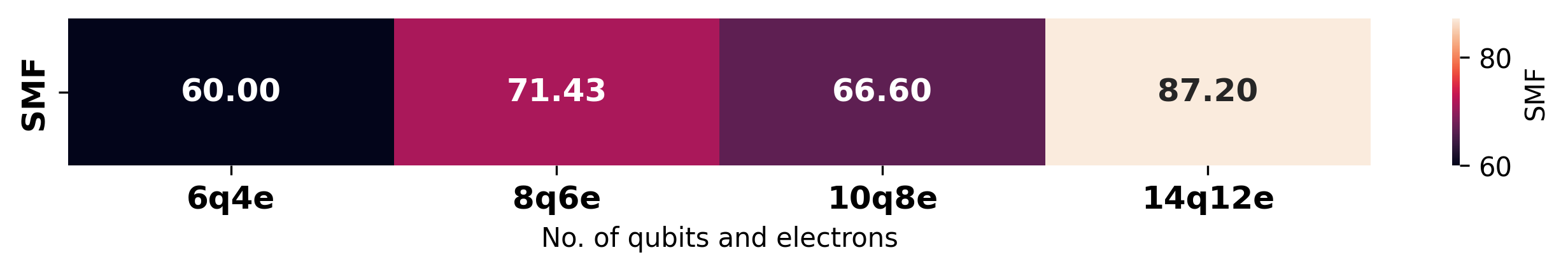}
        \caption{SMF values for HI Molecule}
        \label{fig:hf_prob}
    \end{subfigure}
    \begin{subfigure}{0.8\textwidth}
        \centering
        \includegraphics[width=0.8\textwidth]{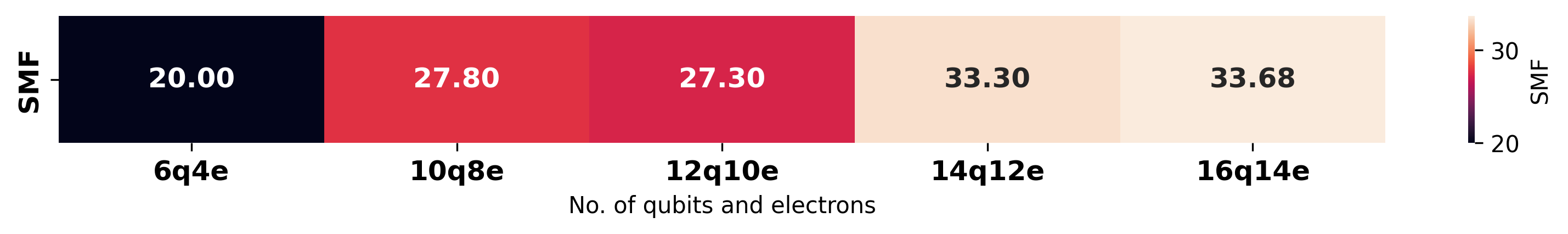}
        \caption{SMF values for \ce{I2} molecule}
        \label{fig:f2_prob}
    \end{subfigure}
    
    \caption{SMF values for reactants and products of Reaction 3.}
    \label{fig:main}
\end{figure*}

\begin{figure*}[h]
    \centering
    \begin{subfigure}{0.8\textwidth}
        \centering
        \includegraphics[width=0.8\textwidth]{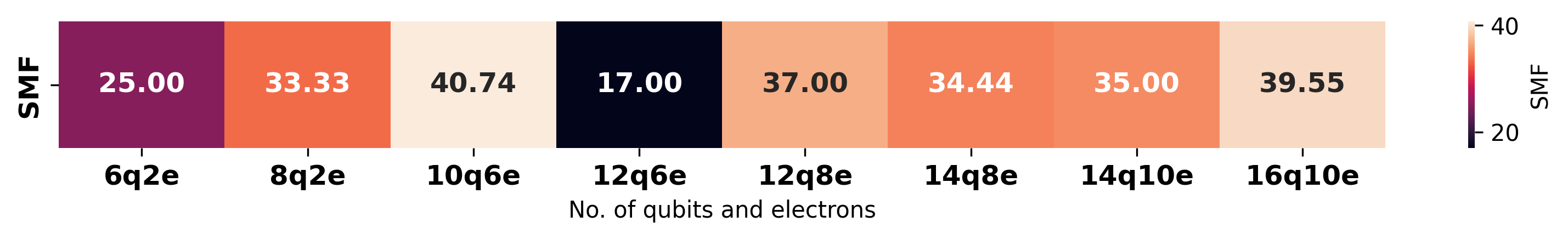}
        \caption{SMF values for CO Molecule}
        \label{fig:hf_prob}
    \end{subfigure}
    \begin{subfigure}{0.8\textwidth}
        \centering
        \includegraphics[width=0.7\textwidth]{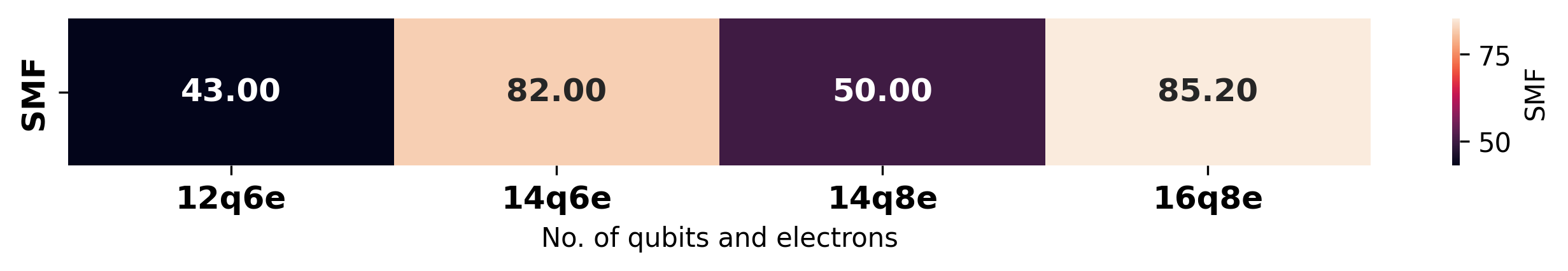}
        \caption{SMF values for \ce{CH4} molecule}
        \label{fig:f2_prob}
    \end{subfigure}
    \begin{subfigure}{0.8\textwidth}
        \centering
        \includegraphics[width=0.8\textwidth]{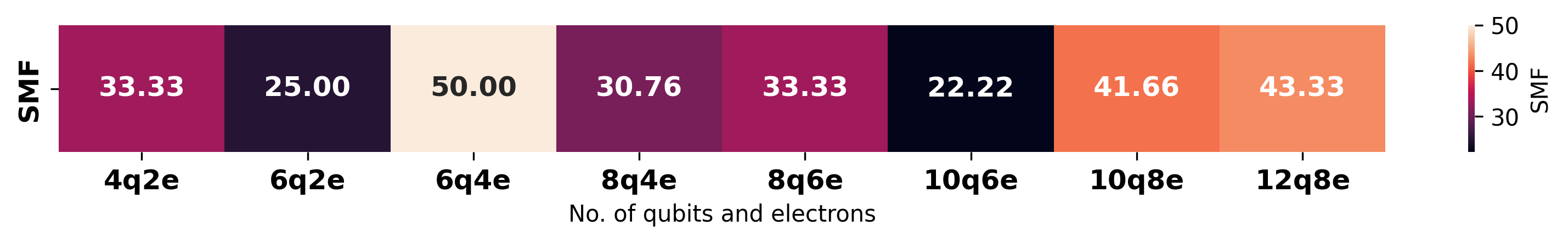}
        \caption{SMF values for \ce{H2O} molecule}
        \label{fig:f2_prob}
    \end{subfigure}
    
    \caption{SMF values for reactants and products of Reaction 4.}
    \label{fig:main}
\end{figure*}

\begin{figure*}[h]
    \centering
    \begin{subfigure}{0.9\textwidth}
        \centering
        \includegraphics[width=0.7\textwidth]{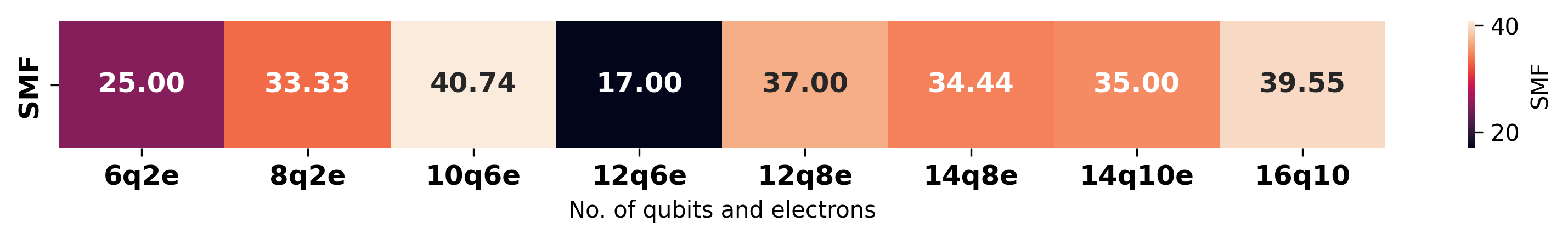}
        \caption{SMF values for \ce{N2} Molecule}
        \label{fig:hf_prob}
    \end{subfigure}
    \begin{subfigure}{0.8\textwidth}
        \centering
        \includegraphics[width=0.8\textwidth]{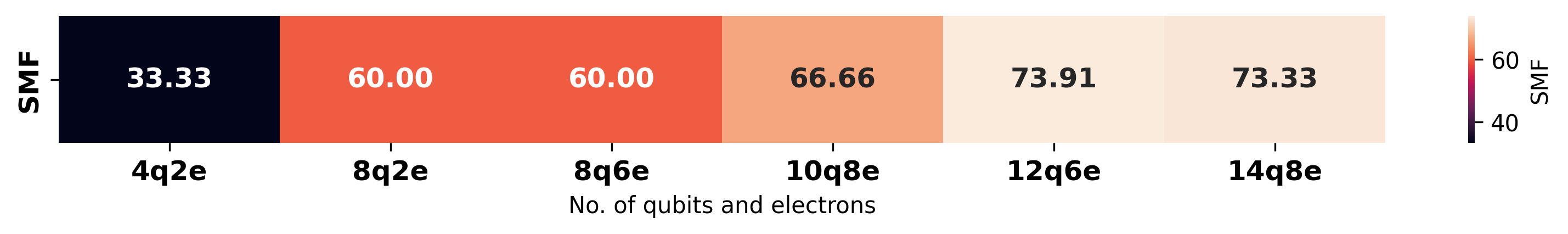}
        \caption{SMF values for \ce{NH3} molecule}
        \label{fig:f2_prob}
    \end{subfigure}
    \caption{SMF values for reactants and products of Reaction 5.}
    \label{fig:main}
\end{figure*}

The HF molecule is a linear diatomic molecule with the point group symmetry $D_{\infty h}$, a non-abelian group. To simplify the calculations, we use the abelian subgroup $C_{2v}$ to determine the irreducible representations of the molecular orbitals and molecular excitations. We find that the irreducible representation for the ground state is $A_1$. Similarly, four of the single excitations and four of the double excitations, out of the total eight excitations, correspond to the irreps $B_2$, $B_2$, $A_2$, $A_2$ and $A_1$, $A_1$, $B_1$, $B_1$, respectively. This indicates that $25\%$  of the excitations share the same irreps as the ground state, as seen in Figure 3(b). Similarly, we calculate the irreducible representations for all the active spaces of reactants and products, and determine how many of them match the ground state. Based on these numbers, we define the SMF. For example, if all the excitations share the same irreducible representation as the ground state, we assign a SMF of $100\%$, and if only half of the excitations match the symmetry of ground state, then the SMF is $50\%$.  \par

As shown in Figure 3(e), for the \ce{F2} molecule, the largest active space considered in our calculations is a 16-qubits or 8 spatial orbitals, 14-electrons system. For the present calculations, two core molecular orbitals were excluded from the active space because the energy separation between these core orbitals and the valence orbitals relevant for chemical bonding is large. Including such deep core orbitals would substantially increase the qubit count and operator complexity in the VQE algorithm while contributing negligibly to the reaction energy of interest. \par

Importantly, the SMF metric does not trivially select the full minimal-basis active space. Only in the case of \ce{H2} does the full minimal basis coincide with a full configuration-interaction active space. For all other molecules studied, the VQE active spaces were selected from a substantially larger set of candidate orbital subsets, and the SMF metric consistently favors proper subspaces rather than the maximal space. For example, in \ce{F2}, the minimal basis contains 10 molecular orbitals, whereas the largest active space used in the SMF-guided VQE calculations comprises only 8 molecular orbitals \par

Figure 3(d) shows that for the \ce{H2} molecule, the 4-qubits or 2 spatial orbitals, 2-electrons method has the symmetry of all excitation terms matching the ground state, resulting in an SMF of $100\%$. In contrast, Figures 3(b) and 3(f) show that for HF (the 10-qubits or 5 spatial orbitals, 8-electrons case) and \ce{F2} (the 16-qubits or 8 spatial orbitals, 14-electrons case), the maximum SMFs are $42\%$ and $36.5\%$, respectively. Clearly, Figure 3(g) demonstrates that the combination of maximum SMFs for all cases achieves chemical accuracy when compared to our CCSD calculations with an accuracy of the order $10^{-5}$, as can be seen in Figure 10, which holds the same comparison for all five reactions.

\subsubsection{\textnormal{\ce{H2 + Cl2} $\rightarrow $ \ce{2HCl}}}

The next reaction studied is the exothermic chlorination of hydrogen. As shown in Figure 5(a), for the HCl molecule, the largest active space used in our calculation is a 10-qubits or 5 spatial orbitals, 8-electrons system. Similarly, for the \ce{Cl2} molecule, as depicted in Figure 5(b), the largest active space is a 16-qubits or 8 spatial orbitals, 14-electrons system. Figures 5(a) and 5(b) also highlight that, for both HCl and \ce{Cl2}, these active spaces correspond to the maximum SMFs— $41.67\%$ and $36.5\%$, respectively. Clearly, Figure 9, which displays the energy values calculated for all the studied reactions, demonstrates that the combination of maximum SMFs achieves chemical accuracy, with an accuracy on the order of $10^{-4}$, out of all 15 possible combinations, as shown in Figure 10. The respective energy values for the different active spaces are available in the supporting information.

\subsubsection{\textnormal{\ce{2HI + Cl2} $\rightarrow$  \ce{2HCl + I2}}}

The third reaction considered is the halogen displacement reaction of chlorine and iodine. As shown in Figure 6(a), for the HI molecule, the largest active space is a 14-qubits or 7 spatial orbitals, 12-electrons system. Similarly, for the \ce{I2} molecule, as illustrated in Figure 6(b), the largest active space is a 16-qubits or 8 spatial orbitals, 14-electrons system. Figures 6(a) and 6(b) also reveal that, for both HI and \ce{I2}, these largest active spaces correspond to the maximum SMFs - $87.2 \%$ and $33.68 \%$, respectively. Figure 9 demonstrates that the combination of maximum SMFs achieves chemical accuracy, with an error on the order of $10^{-1}$, out of all 300 possible combinations, as shown in Figure 10. The respective energy values for the different active spaces can be found in the supporting information.

\subsubsection{\textnormal{\ce{3H2 + CO \rightarrow CH4 + H2O}}}

The fourth reaction studied is the hydrogenation of carbon monoxide, which is known as the Sabatier reaction for the generation of methane from carbon monoxide. Figure 7(a) shows that for the CO molecule, the 16-qubits or 8 spatial orbitals, 10-electrons active space is the largest system considered for the purpose of calculation in this article. Similarly, figures 7(b) and 7(c) show that the largest active space for \ce{CH4} and \ce{H2O} molecules are 16-qubits or 8 spatial orbitals, 8-electrons and 12-qubits or 6 spatial orbitals, 8-electrons systems, respectively. Figures 7(a), 7(b), and 7(c) further demonstrate that for CO, \ce{CH4}, and \ce{H2O}, 10-qubits (5 spatial orbitals), 6-electrons, 16-qubits (8 spatial orbitals) 8-electrons, and 6-qubits (3 spatial orbitals) 4-electrons systems, lead to the maximum SMFs of $40.74\%$, $85.2\%$ and $50\%$, respectively. Figure 9 demonstrates that the combination of maximum SMFs achieves chemical accuracy with an accuracy of the order of $10^{-3}$ out of all possible 256 combinations as shown in Figure 10. Respective energy values for different active spaces are available in the supporting information.

\subsubsection{\textnormal{\ce{3H2 + N2} $\rightarrow$ \ce{2NH3}}}

The final reaction considered is the synthesis of ammonia, which is a crucial step in the Haber process for synthesizing ammonia for fertilizers (nitrogen fixation). As shown in Figure 8(a), for the \ce{N2} molecule, the largest active space is a 16-qubits or 8 spatial orbitals, 10-electrons system. Similarly, for the \ce{NH3} molecule, as depicted in Figure 8(b), the largest active space is a 14-qubits or 7 spatial orbitals, 8-electrons system. Figures 8(a) and 8(b) also indicate that for \ce{N2}  and \ce{NH3}, the 10-qubits (5 spatial orbitals), 6-electrons, and 12-qubits (6 spatial orbitals), 6-electrons systems, the maximum SMFs are $40.74\%$and $73.91\%$, respectively. Figure 9 demonstrates that the combination of these maximum SMFs achieves chemical accuracy with error of the order of $10^{-3}$, out of all $48$ possible combinations, as shown in Figure 10. The respective energy values for the different active spaces are available in the supporting information. \par
Because the symmetry-match fraction (SMF) quantifies the fraction of symmetry-allowed excitation operators within a chosen orbital subset, it enables the a priori screening of active spaces and identifies those capable of supporting the relevant correlation pathways under UCCSD. This significantly reduces resource requirements: SMF-consistent active spaces often achieve chemical accuracy with substantially fewer orbitals and qubits, even for chemically nontrivial systems such as CO. Importantly, this assessment can be made in advance, without the need for empirical testing across many candidate active spaces.

\begin{figure*}[h]
\centering
\begin{subfigure}{0.3\textwidth}
\centering
\includegraphics[width=\linewidth]{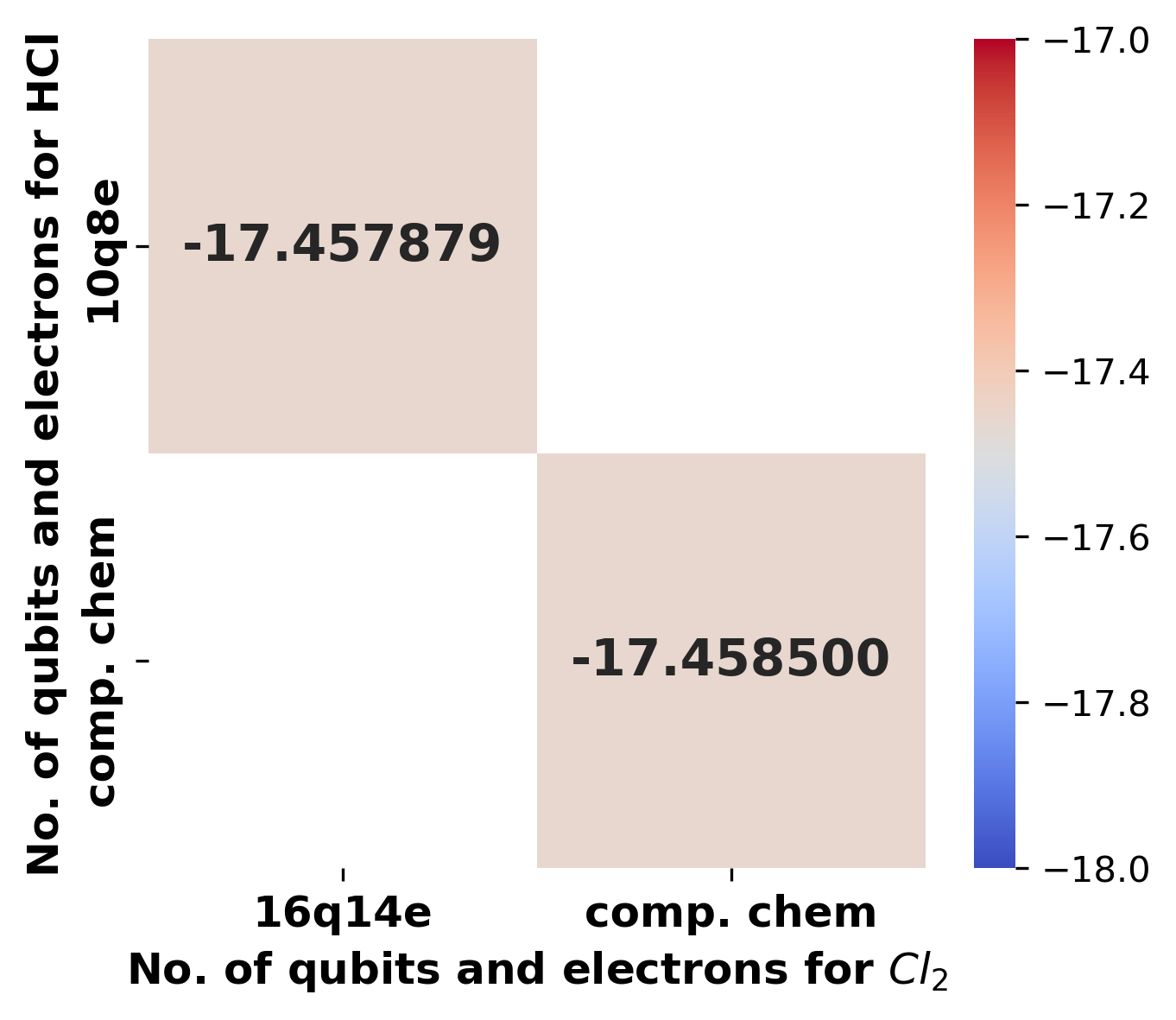} 
\caption{Reaction energy for reaction 2}
\end{subfigure}
\begin{subfigure}{0.3\textwidth}
\centering
\includegraphics[width=\linewidth]{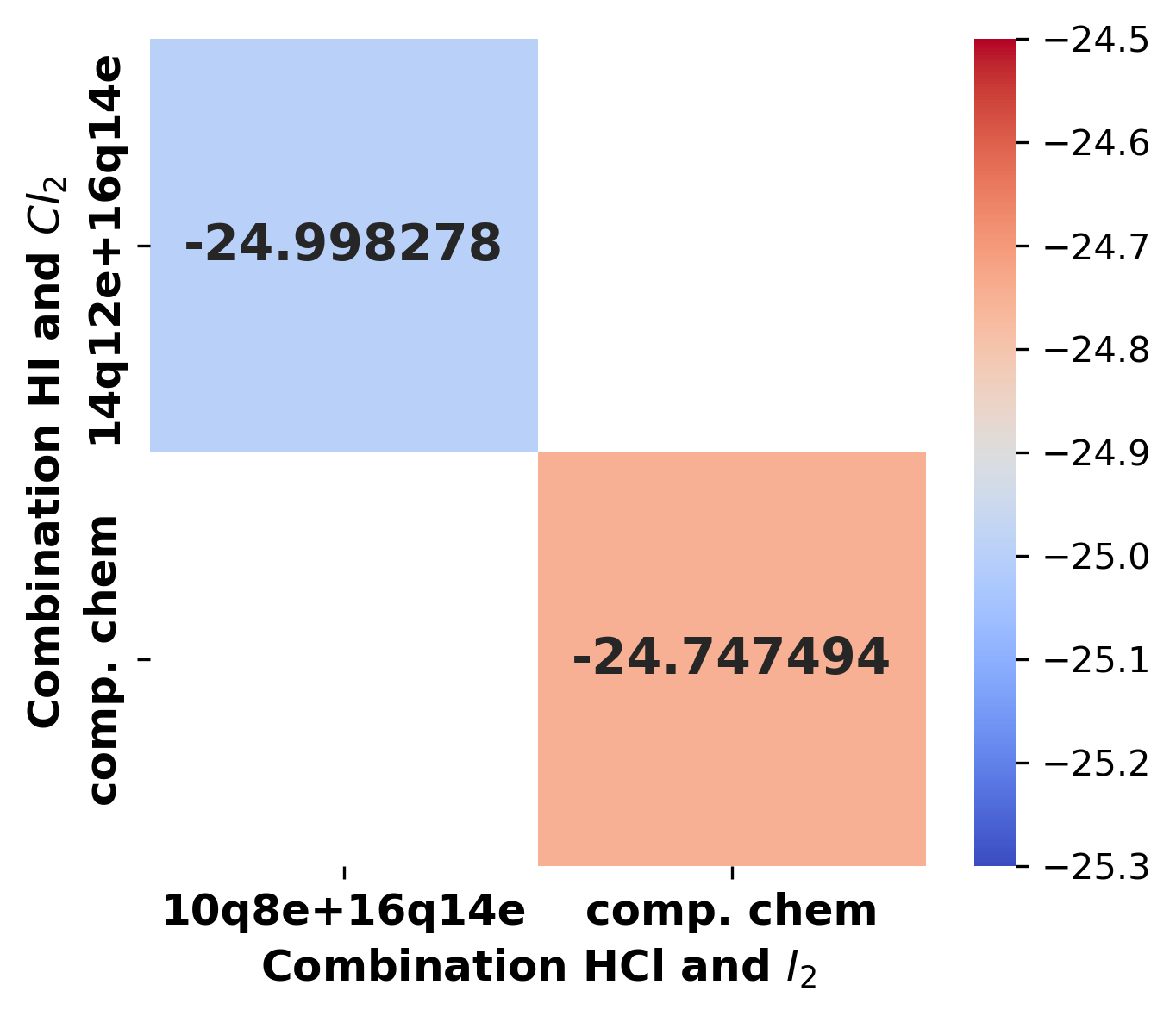}
\caption{Reaction energy for reaction 3}
\label{fig:subim2}
\end{subfigure}
\begin{subfigure}{0.3\textwidth}
\centering
\includegraphics[width=\linewidth]{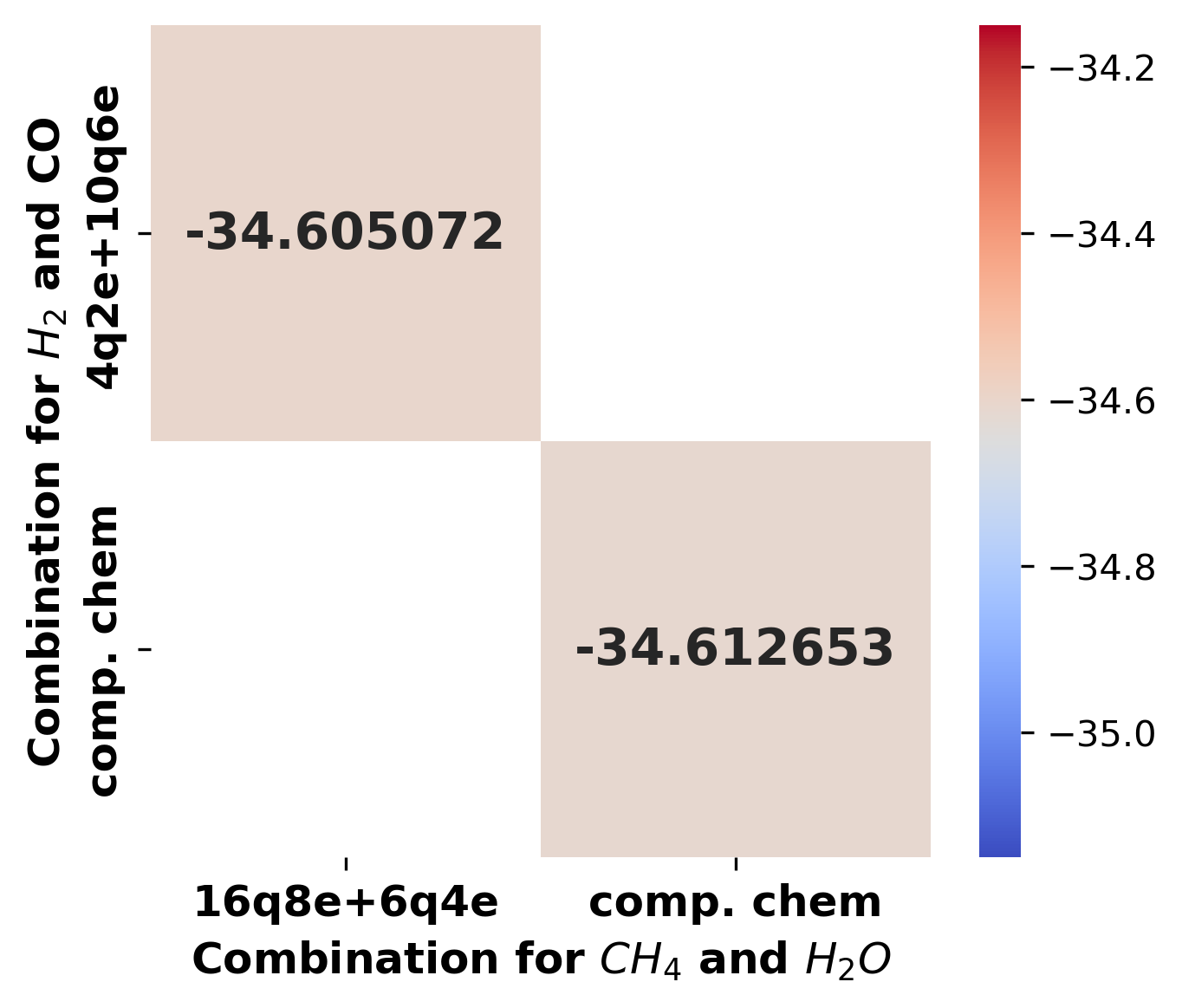}
\caption{Reaction energy for reaction 4}
\label{fig:subim2}
\end{subfigure}
\begin{subfigure}{0.3\textwidth}
\centering
\includegraphics[width=\linewidth]{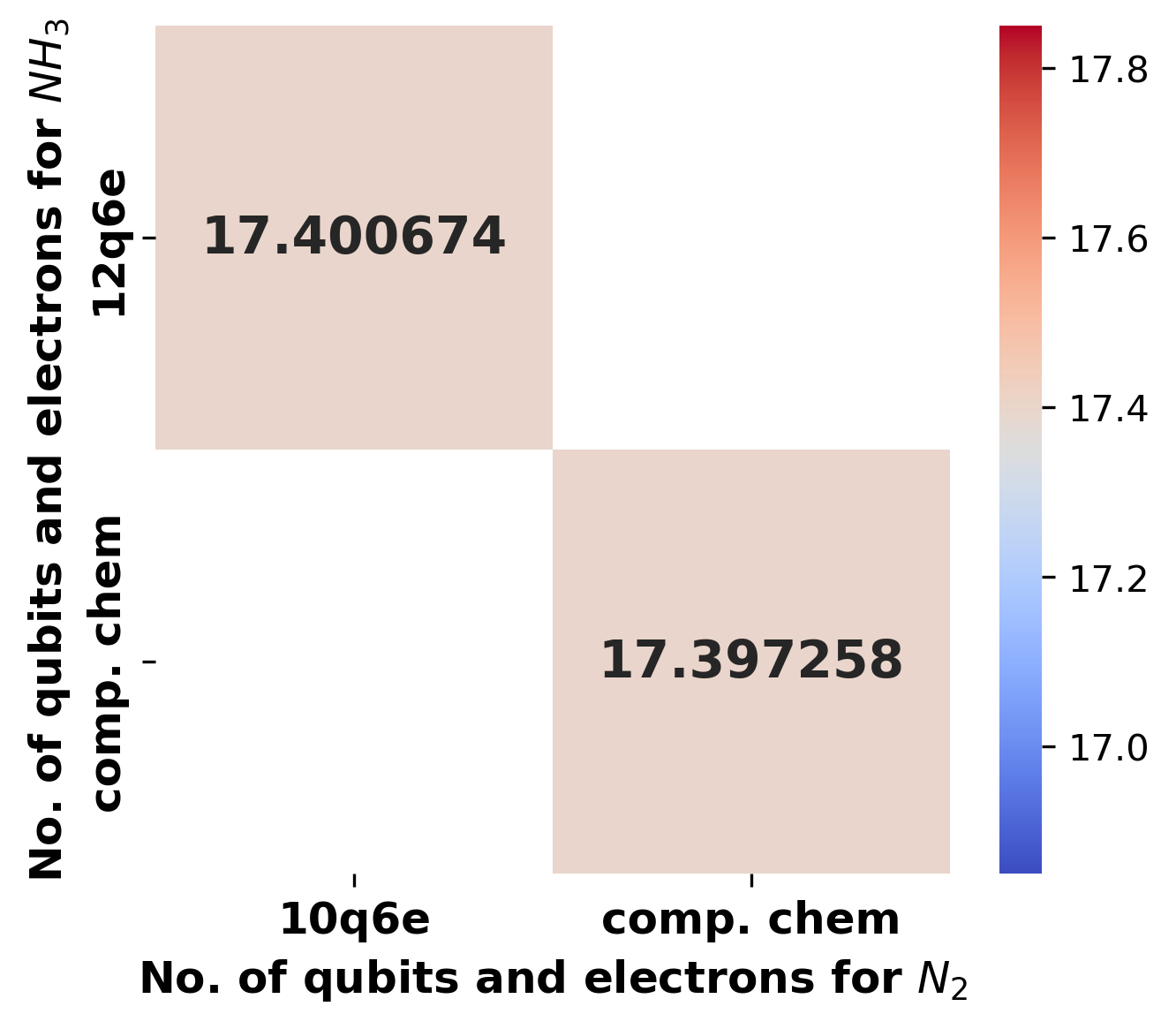} 
\caption{Reaction energy for reaction 5}
\end{subfigure}%
\caption{All reaction energies considering the best combinations}
\label{fig:main}
\end{figure*}

\begin{figure}[h]
\centering
\includegraphics[width=9cm]{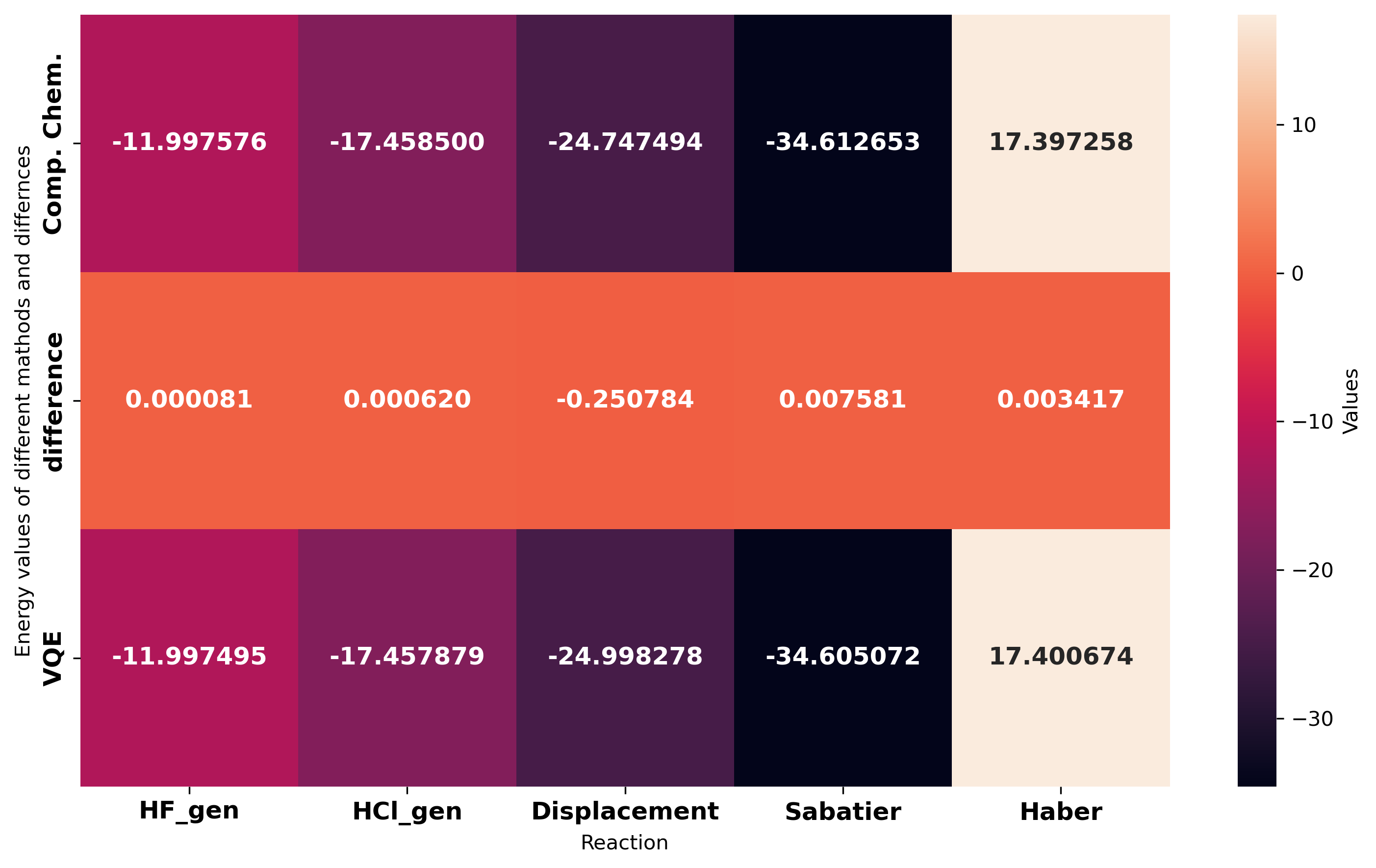}
\caption{Comparison of energy values for all five reactions}
\end{figure}

\subsection{Discussion}
The significance of the findings in the present work in addressing computational chemistry problems in the realm of quantum computation, particularly within the limitations of the NISQ era, can be summarized as follows:

\begin{itemize}
    
\item We have validated the defined scheme for five different reactions, generating expected results for all cases. For all five reactions, the difference in reaction energies between conventional computational chemistry methods and the quantum-classical hybrid VQE algorithm is less than 1 kcal/mol, signifying chemical accuracy, as shown in Figure 10.

\item As the active space size increases, the ground state energy evaluated using our approach converges to the CCSD value. 

\item Our analysis has successfully reduced the number of combinations for each reaction to a single combination. For instance, out of 15 combinations for the first two reactions, 300 combinations for the third reaction, 256 combinations, and 48 combinations for the last reaction, only one combination is needed for each case, as seen in Figure 11.

\item For reactions where most of the molecular orbitals were considered as part of the active space in the calculations (e.g., reactions 1, 2, 4, and 5), the reaction energy difference between the CCSD method and VQE algorithm energies is accurate to or less than two decimal points, as shown in Figure 10. However, for the larger molecule in reaction 3 (involving \ce{I2} and \ce{HI}), when a large portion of the molecular orbitals (MOs) remains uninvolved in the active space, the difference in reaction energy is a bit higher. Nevertheless, it still remains within the limit of chemical accuracy, as can be observed in Figures 9(b) and 10.
 
\end{itemize}

\begin{figure}[h]
\includegraphics[width=9cm]{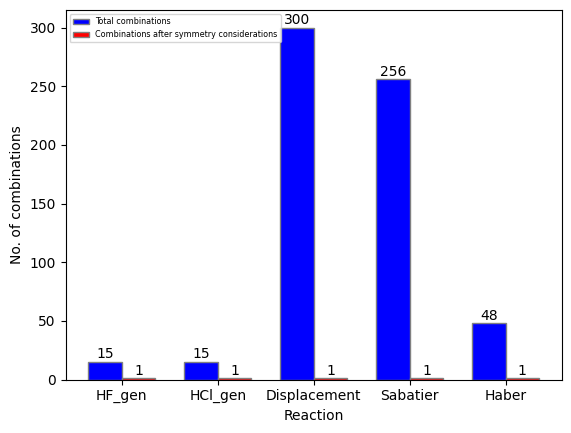}
\caption{Reduction in the number of combinations}
\end{figure}

\section{Conclusion}
Variational algorithms have emerged as a key approach in utilizing NISQ devices for complex optimization problems, operating within a hybrid quantum-classical framework. However, despite their promising potential on NISQ devices, their performance has not met expectations due to several challenges, including noise-sensitive quantum gates, limited qubit connectivity, and the issue of barren plateaus. Therefore, simulating molecules on a quantum computer presents a different set of challenges in the NISQ era, particularly when considering various active spaces for ground state energy (GSE) calculations. As the active spaces increase, the complexity of the calculations also increases. Due to the limited power of NISQ devices, computations involving more than 16 qubits become computationally expensive. \par

To address this, we developed a group-theory-informed framework to identify optimal combinations of qubits (from molecular orbitals) and electrons for different reactants and products. Rather than applying explicit qubit or ansatz parameter reduction, we used irreducible representations (irreps) of the ground and excited states to quantify the symmetry-consistent excitations that contribute most significantly to the ground state energy (GSE). Based on this, we defined a symmetry-matched fraction (SMF) measure for excitation relevance and selected, for each species, the active space with the highest symmetry-aligned excitation SMF. This method enables us to maintain chemical accuracy, comparable to traditional quantum chemistry techniques. We validated this scheme across five chemically and industrially relevant reactions and observed consistent and accurate results in each case. Notably, this method reduced the total number of active space combinations from potentially hundreds to a single optimal configuration per reaction.  For instance, as shown in Figure 11, the number of considered combinations decreased from $15$ to $1$ for the first two reactions, from $300$ to $1$ for the third, from $256$ to $1$ for the fourth, and from $48$ to $1$ for the fifth. Moreover, the reaction energy differences remained well within the threshold of 1 kcal/mol for all cases (Figure 10), with very good agreement between CCSD and VQE energies for smaller systems. Even for larger molecules, the accuracy was preserved within chemical accuracy. \par
Therefore, by extending the use of symmetry beyond circuit simplification, our work complements existing strategies and contributes to the broader goal of making quantum simulations more predictive and scalable. In contrast to molecule-specific demonstrations or hardware-focused implementations, we offer a generalizable methodology applicable across diverse reaction classes. Designed with the constraints of current NISQ hardware in mind, our framework provides a practical path toward quantum-enabled reaction modeling, helping bridge the gap between algorithmic development and real-world chemical applications. In summary, the SMF metric does not replace existing symmetry-preserving ansätze or adaptive algorithms; rather, it complements them by enabling the a priori identification of compact, symmetry-consistent active spaces prior to circuit construction. \par 
As part of our future work, we will investigate how this scheme performs with larger basis sets and whether chemical accuracy can still be achieved in such scenarios, as the molecular orbitals become significantly complicated. Additionally, we plan to explore the use of symmetry methods to simplify the Hamiltonian and wave functions in order to simulate different reactions and reduce computational costs. We will also examine the impact of noise on the results and explore error mitigation techniques that leverage symmetry relations to achieve chemical accuracy in quantum chemistry problems. \par

\begin{acknowledgments}
MS, LR, APG, AK, and MP gratefully acknowledge the Indian Institute of Technology, Jodhpur, for providing the computational facilities essential for completing this work. AK also acknowledges SERB (Grant No. CRG/2022/005979) for funding the project. MS further thanks the Department of Chemistry, IIT Jodhpur, and the Ministry of Education (MoE) for offering research infrastructure and financial support.
\end{acknowledgments}





\bibliography{apssamp}
\end{document}



\title{\Huge\bfseries Quantum Simulations of Chemical Reactions: Achieving Accuracy with NISQ Devices:\\
Supporting Information} 

\author{Maitreyee Sarkar$^{1}$}
\author{Lisa Roy$^{1}$}%
 \author{Akash Gutal$^{2}$}%
 \author{Atul Kumar$^{1}$}%
 \email{atulk@iitj.ac.in}
 \author{Manikandan Paranjothy$^{2}$}%
 \email{pmanikandan@iitj.ac.in}
\affiliation{%
$^{1}$ Quantum Information and Computation Lab, Department of Chemistry,
 Indian Institute of Technology Jodhpur, Rajasthan, India, 342030 
}%
\affiliation{%
$^{2}$ Chemical Dynamics Research Group, Department of Chemistry,
 Indian Institute of Technology Jodhpur, Rajasthan, India, 342030 
}%

\maketitle


\section{Optimized Geometries}
Geometry optimization values for all reactants and products, calculated using CCSD/STO-3G with NWChem, where the coordinates are expressed in Angstrom units.

\begin{figure}[h]
\centering
\begin{subfigure}{0.3\textwidth}
\centering
\includegraphics[width=3cm]{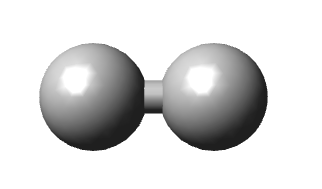} 
\end{subfigure}
\hspace{0.05\textwidth} 
\begin{subfigure}{0.3\textwidth}
\centering
\includegraphics[height=3cm]{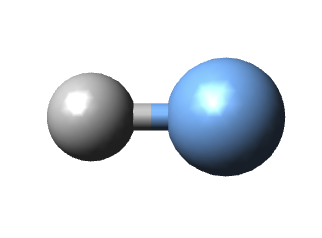}
\end{subfigure}
\end{figure}

\begin{table}[h]
\begin{tabular}{ccccccccccc}
   H      & 0.00000000  &   0.00000000  &   -0.36743273 & &&&F & 0.00000000    & 0.00000000   &   -0.12163802 \\
   H      & 0.00000000 &    0.00000000  &   0.36743273 & &&&H & 0.00000000    & 0.00000000  &   0.87299002\\
\end{tabular}
\end{table}


\begin{figure}[h!]
\centering
\begin{subfigure}{0.3\textwidth}
\centering
\includegraphics[width=3cm]{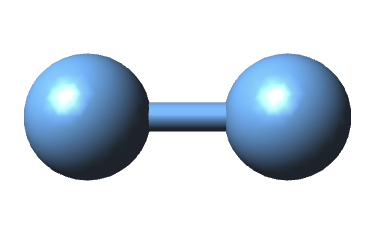} 
\end{subfigure}
 \hspace{0.05\textwidth} 
\begin{subfigure}{0.3\textwidth}
\centering
\includegraphics[width=4cm]{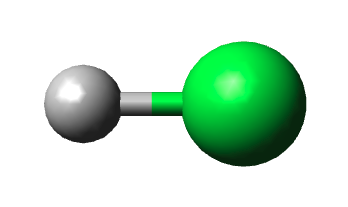}
\end{subfigure}
\end{figure}

\begin{table}[h!]
\begin{tabular}{ccccccccccc}
   F & 0.00000000   &  0.00000000   & -0.69382832 & &&&Cl & 0.00000000 &     0.00000000   &  -0.09087167 \\
   F & 0.00000000   &  0.00000000   &  0.69382832 & &&&H  & 0.00000000   &   0.00000000 &     1.25160055\\
\end{tabular}
\end{table}


\begin{figure}[h]
\centering
\begin{subfigure}{0.3\textwidth}
\centering
\includegraphics[width=3cm]{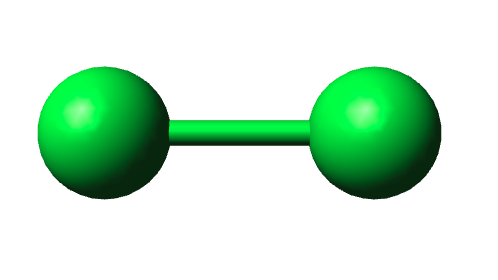} 
\end{subfigure}
 \hspace{0.05\textwidth} 
\begin{subfigure}{0.3\textwidth}
\centering
\includegraphics[width=4cm]{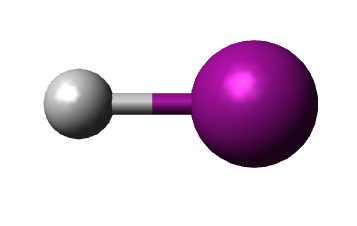}
\end{subfigure}
\end{figure}

\begin{table}[h]
\begin{tabular}{ccccccccccc}
   Cl &   0.00000000   &   0.00000000&     -1.06485007  & &&&I           &         0.00000000   &   0.00000000 &    -0.01919551 \\
   Cl   & 0.00000000 &     0.00000000   &   1.06485007 & &&&H             &       0.00000000 &     0.00000000   &   1.60993328\\
\end{tabular}
\end{table}


\begin{figure}[h]
\centering
\begin{subfigure}{0.3\textwidth}
\centering
\includegraphics[width=4.5cm]{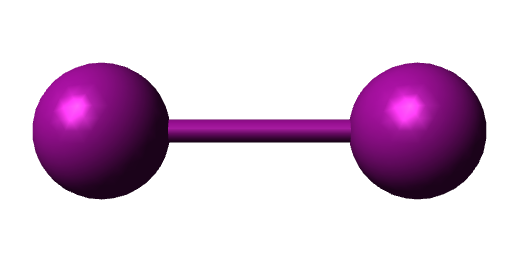} 
\end{subfigure}
 \hspace{0.05\textwidth} 
\begin{subfigure}{0.3\textwidth}
\centering
\includegraphics[width=2.5cm]{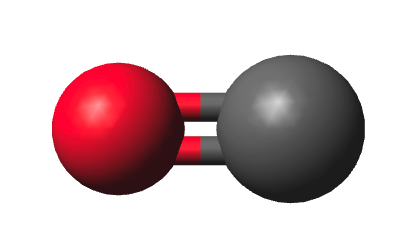}
\end{subfigure}
\end{figure}

\begin{table}[h]
\begin{tabular}{ccccccccccc}
   I    &     0.00000000   &   0.00000000  &   -1.36900912 & &&&C&        0.00000000&      0.00000000 &    -0.66665088 \\
   I      &   0.00000000 &     0.00000000    &  1.36900912 & &&&O  &      0.00000000  &    0.00000000   &   0.51565030\\
\end{tabular}
\end{table}

\begin{figure}[h]
\centering
\begin{subfigure}{0.3\textwidth}
\centering
\includegraphics[width=3cm]{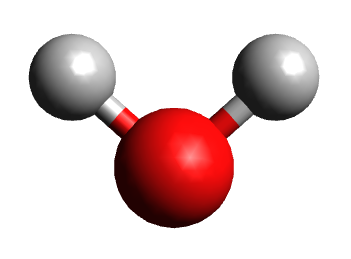} 
\end{subfigure}
 \hspace{0.05\textwidth} 
\begin{subfigure}{0.3\textwidth}
\centering
\includegraphics[width=3cm]{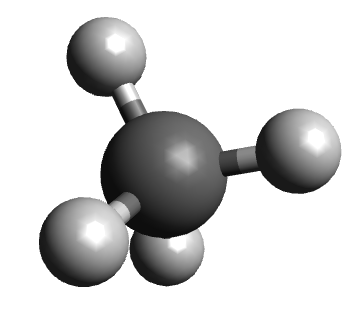}
\end{subfigure}
\end{figure}

\begin{table}[h]
\begin{tabular}{ccccccccccc}
   O  &  0.00000000     & 0.00000000 &    -0.14920525 & &&& C&        -0.00004309&     -0.00000945&      0.00002230  \\
   H  & -0.76875446    &  0.00000000   &   0.53372741 & &&&H  &       1.10818228  &   -0.00001016  &   -0.00001137\\
    H   & 0.76875446  &    0.00000000     & 0.53372741 & &&&H&-0.36936923    & -0.78861297    &  0.68544214\\
         &    &        &   & &&&H       & -0.36939931     & 0.98789545     & 0.34017445 \\
              &    &        &   & &&&H       & -0.36939931     & 0.98789545     & 0.34017445 \\
\end{tabular}
\end{table}


\begin{figure}[h]
\centering
\begin{subfigure}{0.3\textwidth}
\centering
\includegraphics[width=2cm]{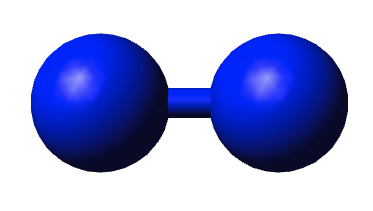} 
\end{subfigure}
 \hspace{0.05\textwidth} 
\begin{subfigure}{0.3\textwidth}
\centering
\includegraphics[width=4cm]{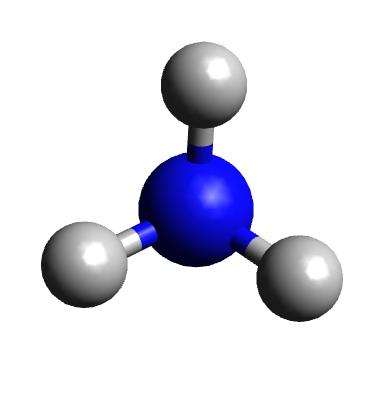}
\end{subfigure}
\end{figure}

\begin{table}[h]
\begin{tabular}{ccccccccccc}
   N  &      0.00000000  &    0.00000000   &  -0.59453116 & &&& N&         0.00000000    & -0.00000000&     -0.19630170  \\
   N    &    0.00000000    &  0.00000000 &     0.59453116 & &&&H  &      -0.52537840    & -0.78820401  &    0.30191059\\
              &    &        &   & &&&H    &     0.94529389  &   -0.06088904    &  0.30191059 \\
              &    &        &   & &&&H      &  -0.41991549&      0.84909305     & 0.30191059 \\
\end{tabular}
\end{table}

\FloatBarrier

\section{Energy values for reactants and products for rest of the four reactions}

\begin{figure*}[h]
    \centering

    \begin{subfigure}{0.8\textwidth}
        \centering
        \includegraphics[width=0.7\textwidth]{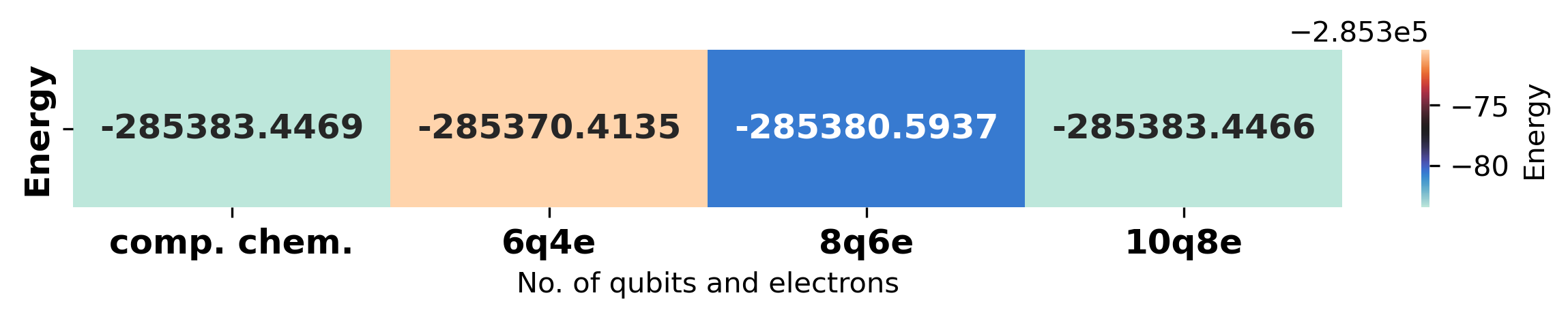}
        \caption{Energy values (in kcal/mol) for HCl Molecule}
        \label{fig:hf_energy}
    \end{subfigure}


    \begin{subfigure}{0.8\textwidth}
        \centering
        \includegraphics[width=0.9\textwidth]{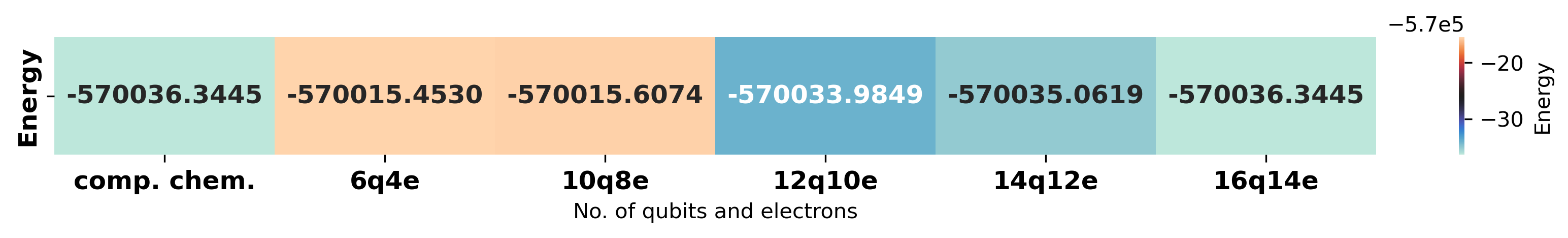}
        \caption{Energy values (in kcal/mol) for \ce{Cl2} molecule}
        \label{fig:f2_energy}
    \end{subfigure}

    
    \caption{\textbf{Energy values (in kcal/mol) for reactants and products of Reaction 2.}}
    \label{fig:main}
\end{figure*}

\begin{figure*}[h]
    \centering

    \begin{subfigure}{0.8\textwidth}
        \centering
        \includegraphics[width=\textwidth]{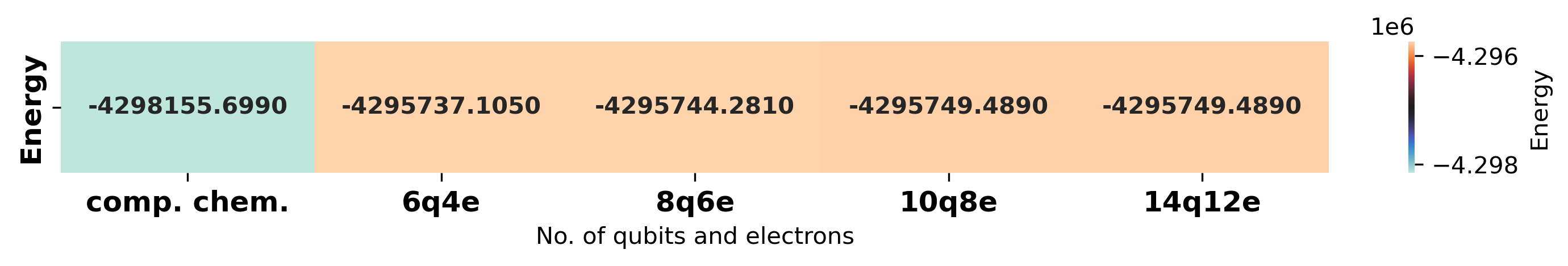}
        \caption{Energy values (in kcal/mol) for HI Molecule}
        \label{fig:hf_energy}
    \end{subfigure}


    \begin{subfigure}{0.8\textwidth}
        \centering
        \includegraphics[width=1.1\textwidth]{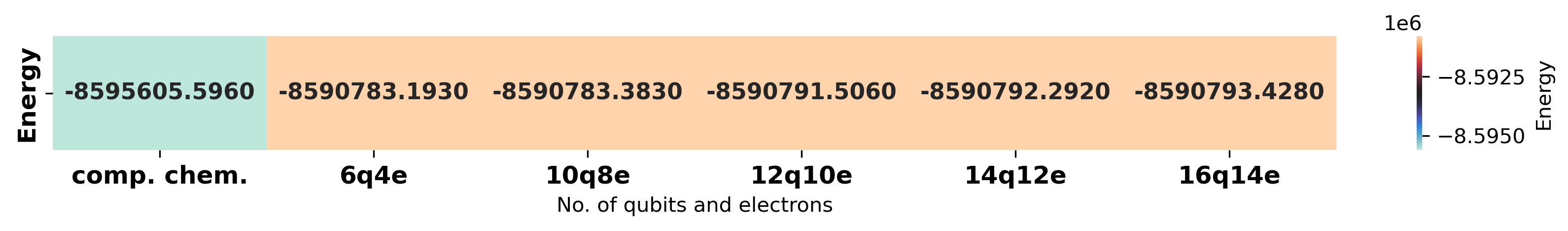}
        \caption{Energy values (in kcal/mol) for \ce{I2} molecule}
        \label{fig:f2_energy}
    \end{subfigure}

    
    \caption{\textbf{Energy values (in kcal/mol) for reactants and products of Reaction 3.}}
    \label{fig:main}
\end{figure*}

\begin{figure*}[h]
    \centering

    \begin{subfigure}{0.8\textwidth}
        \centering
        \includegraphics[width=1.1\textwidth]{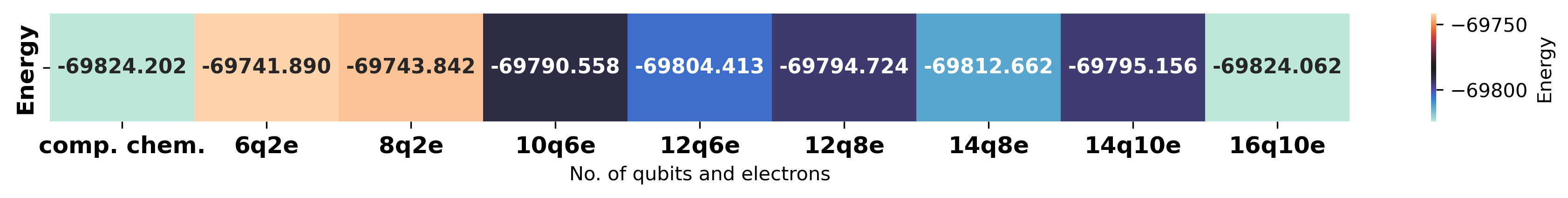}
        \caption{Energy values (in kcal/mol) for CO Molecule}
        \label{fig:hf_energy}
    \end{subfigure}
    \begin{subfigure}{0.8\textwidth}
        \centering
        \includegraphics[width=0.9\textwidth]{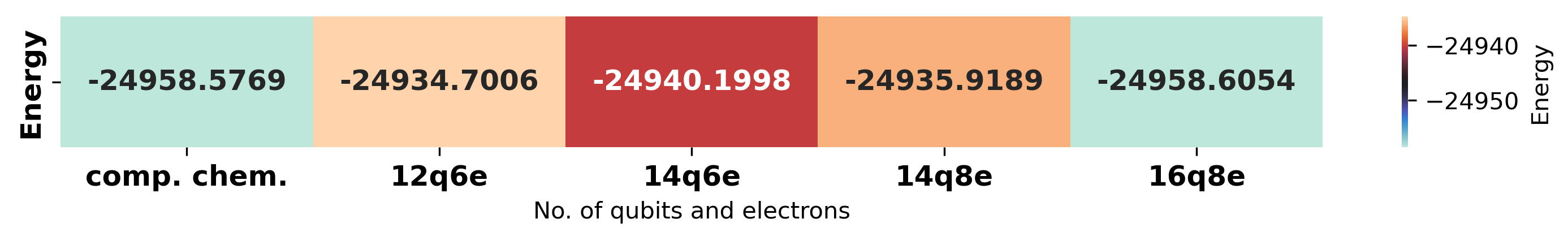}
        \caption{Energy values (in kcal/mol) for \ce{CH4} molecule}
        \label{fig:f2_energy}
    \end{subfigure}
    \begin{subfigure}{0.8\textwidth}
        \centering
        \includegraphics[width=1.1\textwidth]{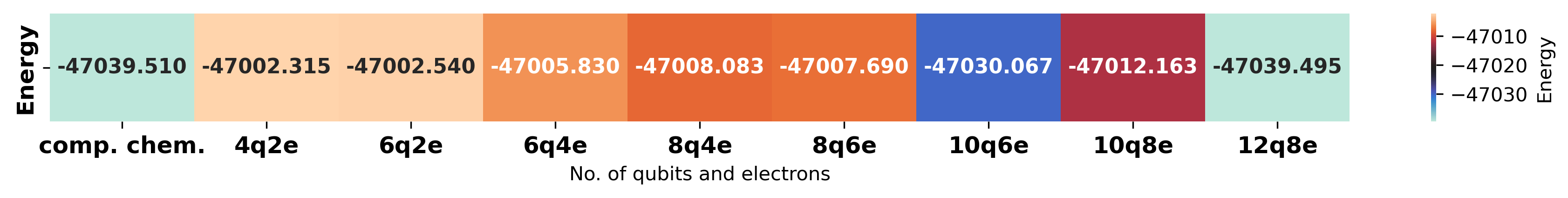}
        \caption{Energy values (in kcal/mol) for \ce{H2O} molecule}
        \label{fig:f2_energy}
    \end{subfigure}
    
    \caption{\textbf{Energy values (in kcal/mol) for reactants and products of Reaction 4.}}
    \label{fig:main}
\end{figure*}

\begin{figure*}[h]
    \centering

    \begin{subfigure}{0.8\textwidth}
        \centering
        \includegraphics[width=1.1\textwidth]{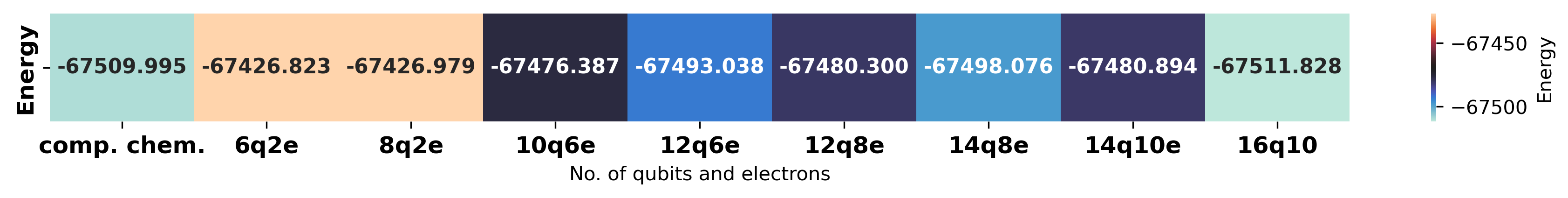}
        \caption{Energy values (in kcal/mol) for \ce{N2} Molecule}
        \label{fig:hf_energy}
    \end{subfigure}
    \begin{subfigure}{0.8\textwidth}
        \centering
        \includegraphics[width=\textwidth]{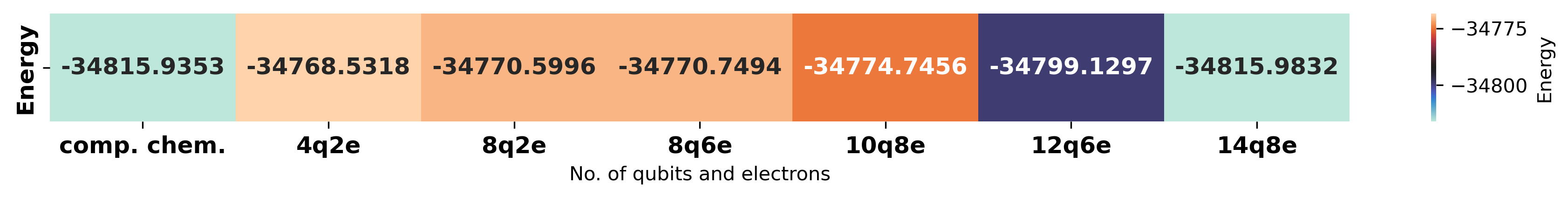}
        \caption{Energy values (in kcal/mol) for \ce{NH3} molecule}
        \label{fig:f2_energy}
    \end{subfigure}
    
    \caption{\textbf{Energy values (in kcal/mol) for reactants and products of Reaction 5.}}
    \label{fig:main}
\end{figure*}

\FloatBarrier
\clearpage


\section{Symmetry of point group}

\begin{table*}[h]
\begin{tabular}{ccccc}\toprule
\\
 \multirow{1}{2cm}{Molecule} & \multirow{1}{2cm}{Point Group} & \multirow{1}{2.5cm}{Group type} & \multirow{1}{4.5cm}{Abelian Subgroup Used}  \\
 \\
 \hline
 \\
\ce{H2} & $D_{\infty h}$ & Non Abelian & $C_{2v}$  \\\\
\ce{HF} & $C_{\infty v}$ & Non Abelian & $C_{2v}$  \\\\
\ce{F2} & $D_{\infty h}$ & Non Abelian & $C_{2v}$  \\\\
\ce{HCl} & $C_{\infty v}$ & Non Abelian & $C_{2v}$  \\\\
\ce{Cl2} & $D_{\infty h}$ & Non Abelian & $C_{2v}$  \\\\
\ce{HI} & $C_{\infty v}$ & Non Abelian & $C_{2v}$  \\\\
\ce{I2} & $D_{\infty h}$ & Non Abelian & $C_{2v}$  \\\\
\ce{CO} & $C_{\infty v}$ & Non Abelian & $C_{2v}$  \\\\
\ce{H2O} & $C_{2v}$ & Abelian & -  \\\\
\ce{CH4} & $T_d$ & Non Abelian & $C_s$   \\\\
\ce{N2} & $D_{\infty h}$ & Non Abelian & $C_{2v}$  \\\\
\ce{NH3} & $C_{3v}$ & Non Abelian & $C_s$  \\\\
 \bottomrule
\end{tabular}
\end{table*}

\FloatBarrier
\clearpage

\section{Molecular Orbitals}
Molecular orbitals for all reactants and products that contributed to the active space selection in VQE, calculated using CCSD/STO-3G with Orca.

\subsection{Orbital Energies}

\begin{table*}[h]
\caption{\label{tab:table4}HF}
\begin{tabular}{ccc}\toprule
\\
 \multirow{1}{1cm}{MO} & \multirow{1}{3cm}{No. of electrons} & \multirow{1}{3cm}{Orbital Energy ($E_h$)} \\
 \\
 \hline
 \\
0 & 2 & -25.907079  \\ \\
1 & 2 & -1.450018  \\ \\
2 & 2 & -0.561731  \\ \\
3 & 2 & -0.462447 \\ \\
4 & 2 & -0.462447 \\ \\
5 & 0 & 0.552308 \\ \\
 \bottomrule
\end{tabular}
\end{table*}

\begin{table*}[h]
\caption{\label{tab:table4}\ce{F2}}
\begin{tabular}{ccc}\toprule
\\
 \multirow{1}{1cm}{MO} & \multirow{1}{3cm}{No. of electrons} & \multirow{1}{3cm}{Orbital Energy ($E_h$)} \\
 \\
 \hline
 \\
0 & 2 & -26.045178 \\ \\
1 & 2 & -26.044173 \\ \\
2 & 2 & -1.66929 \\ \\
3 & 2 & -1.331813 \\ \\
4 & 2 & -0.633172 \\ \\
5 & 2 & -0.633172 \\ \\
6 & 2 & -0.586152 \\ \\
7 & 2 & -0.458071 \\ \\
8 & 2 & -0.458071 \\ \\
9 & 0 & 0.421907 \\ \\
 \bottomrule
\end{tabular}
\end{table*}

\begin{table*}[h]
\caption{\label{tab:table4}\ce{H2O}}
\begin{tabular}{ccc}\toprule
\\
 \multirow{1}{1cm}{MO} & \multirow{1}{3cm}{No. of electrons} & \multirow{1}{3cm}{Orbital Energy ($E_h$)} \\
 \\
 \hline
 \\
0 & 2 & -20.244398 \\ \\
1 & 2 & -1.27037 \\ \\
2 & 2 & -0.614402 \\ \\
3 & 2 & -0.457504 \\ \\
4 & 2 & -0.392621 \\ \\
5 & 0 & 0.608334 \\ \\
6 & 0 & 0.736142 \\ \\
 \bottomrule
\end{tabular}
\end{table*}

\begin{table*}[h]
\caption{\label{tab:table4}\ce{CH4}}
\begin{tabular}{ccc}\toprule
\\
 \multirow{1}{1cm}{MO} & \multirow{1}{3cm}{No. of electrons} & \multirow{1}{3cm}{Orbital Energy ($E_h$)} \\
 \\
 \hline
 \\
0 & 2 & -11.03007 \\ \\
1 & 2 & -0.9098 \\ \\
2 & 2 & -0.518757 \\ \\
3 & 2 & -0.518744 \\ \\
4 & 2 & -0.518727 \\ \\
5 & 0 & 0.715244 \\ \\
6 & 0 & 0.715284 \\ \\
7 & 0 & 0.715311 \\ \\
8 & 0 & 0.754338 \\ \\
 \bottomrule
\end{tabular}
\end{table*}

\begin{table*}[h]
\caption{\label{tab:table4}CO}
\begin{tabular}{ccc}\toprule
\\
 \multirow{1}{1cm}{MO} & \multirow{1}{3cm}{No. of electrons} & \multirow{1}{3cm}{Orbital Energy ($E_h$)} \\
 \\
 \hline
 \\
0 & 2 & -20.41723 \\ \\
1 & 2 & -11.092398 \\ \\
2 & 2 & -1.448081 \\ \\
3 & 2 & -0.697329 \\ \\
4 & 2 & -0.542059 \\ \\
5 & 2 & -0.542059 \\ \\
6 & 2 & -0.445393 \\ \\
7 & 0 & 0.307528 \\ \\
8 & 0 & 0.307528 \\ \\
9 & 0 & 1.015011 \\ \\
 \bottomrule
\end{tabular}
\end{table*}

\begin{table*}[h]
\caption{\label{tab:table4}\ce{N2}}
\begin{tabular}{ccc}\toprule
\\
 \multirow{1}{1cm}{MO} & \multirow{1}{3cm}{No. of electrons} & \multirow{1}{3cm}{Orbital Energy ($E_h$)} \\
 \\
 \hline
 \\
0 & 2 & -15.512675 \\ \\
1 & 2 & -15.511009 \\ \\
2 & 2 & -1.42746 \\ \\
3 & 2 & -0.724705 \\ \\
4 & 2 & -0.562084 \\ \\
5 & 2 & -0.562084 \\ \\
6 & 2 & -0.535386 \\ \\
7 & 0 & 0.274106 \\ \\
8 & 0 & 0.274106 \\ \\
9 & 0 & 1.085938 \\ \\
 \bottomrule
\end{tabular}
\end{table*}

\begin{table*}[h]
\caption{\label{tab:table4}\ce{NH3}}
\begin{tabular}{ccc}\toprule
\\
 \multirow{1}{1cm}{MO} & \multirow{1}{3cm}{No. of electrons} & \multirow{1}{3cm}{Orbital Energy ($E_h$)} \\
 \\
 \hline
 \\
0 & 2 & -15.306729 \\ \\
1 & 2 & -1.091891 \\ \\
2 & 2 & -0.571494 \\ \\
3 & 2 & -0.571494 \\ \\
4 & 2 & -0.354679 \\ \\
5 & 0 & 0.640769 \\ \\
6 & 0 & 0.72617 \\ \\
7 & 0 & 0.72617 \\ \\
 \bottomrule
\end{tabular}
\end{table*}

\begin{table*}[h]
\caption{\label{tab:table4}\ce{HCl}}
\begin{tabular}{ccc}\toprule
\\
 \multirow{1}{1cm}{MO} & \multirow{1}{3cm}{No. of electrons} & \multirow{1}{3cm}{Orbital Energy ($E_h$)} \\
 \\
 \hline
 \\
0 & 2 & -103.722791 \\ \\
1 & 2 & -10.40429 \\ \\
2 & 2 & -7.849161 \\ \\
3 & 2 & -7.842429 \\ \\
4 & 2 & -7.842429 \\ \\
5 & 2 & -1.044291 \\ \\
6 & 2 & -0.565075 \\ \\
7 & 2 & -0.423358 \\ \\
8 & 2 & -0.423358 \\ \\
9 & 0 & 0.405831 \\ \\
 \bottomrule
\end{tabular}
\end{table*}

\begin{table*}[h]
\caption{\label{tab:table4}\ce{Cl2}}
\begin{tabular}{ccc}\toprule
\\
 \multirow{1}{1cm}{MO} & \multirow{1}{3cm}{No. of electrons} & \multirow{1}{3cm}{Orbital Energy ($E_h$)} \\
 \\
 \hline
 \\
0 & 2 & -103.795634 \\ \\
1 & 2 & -103.795583 \\ \\
2 & 2 & -10.473339 \\ \\
3 & 2 & -10.471206 \\ \\
4 & 2 & -7.918623 \\ \\
5 & 2 & -7.916281 \\ \\
6 & 2 & -7.91029 \\ \\
7 & 2 & -7.91029 \\ \\
8 & 2 & -7.909448 \\ \\
9 & 2 & -7.909448 \\ \\
10 & 2 & -1.145444 \\ \\
11 & 2 & -0.942027 \\ \\
12 & 2 & -0.526112 \\ \\
13 & 2 & -0.521324 \\ \\
14 & 2 & -0.521324 \\ \\
15 & 2 & -0.409464 \\ \\
16 & 2 & -0.409464 \\ \\
17 & 0 & 0.116462 \\ \\
 \bottomrule
\end{tabular}
\end{table*}

\begin{table*}[h]
\caption{\label{tab:table4}HI}
\begin{tabular}{ccc}\toprule
\\
 \multirow{1}{1cm}{MO} & \multirow{1}{3cm}{No. of electrons} & \multirow{1}{3cm}{Orbital Energy ($E_h$)} \\
 \\
 \hline
 \\
12 & 2 & -23.374046 \\ \\
13 & 2 & -23.374046 \\ \\
14 & 2 & -6.56669 \\ \\
15 & 2 & -5.178154 \\ \\
16 & 2 & -5.161543 \\ \\
17 & 2 & -5.161543 \\ \\
18 & 2 & -1.984738 \\ \\
19 & 2 & -1.97773 \\ \\
20 & 2 & -1.97773 \\ \\
21 & 2 & -1.96336 \\ \\
22 & 2 & -1.96336 \\ \\
23 & 2 & -0.730611 \\ \\
24 & 2 & -0.434482 \\ \\
25 & 2 & -0.293176 \\ \\
26 & 2 & -0.293176 \\ \\
27 & 0 & 0.343146 \\ \\
 \bottomrule
\end{tabular}
\end{table*}

\begin{table*}[h]
\caption{\label{tab:table4}\ce{I2}}
\begin{tabular}{ccc}\toprule
\\
 \multirow{1}{1cm}{MO} & \multirow{1}{3cm}{No. of electrons} & \multirow{1}{3cm}{Orbital Energy ($E_h$)} \\
 \\
 \hline
 \\
24 & 2 & -23.397506 \\ \\
25 & 2 & -23.397506 \\ \\
26 & 2 & -23.397506 \\ \\
27 & 2 & -23.397506 \\ \\
28 & 2 & -6.591794 \\ \\
29 & 2 & -6.591725 \\ \\
30 & 2 & -5.20538 \\ \\
31 & 2 & -5.204968 \\ \\
32 & 2 & -5.185372 \\ \\
33 & 2 & -5.185372 \\ \\
34 & 2 & -5.185338 \\ \\
35 & 2 & -5.185338 \\ \\
36 & 2 & -2.009892 \\ \\
37 & 2 & -2.009729 \\ \\
38 & 2 & -2.003544 \\ \\
39 & 2 & -2.003544 \\ \\
40 & 2 & -2.00346 \\ \\
41 & 2 & -2.00346 \\ \\
42 & 2 & -1.986513 \\ \\
43 & 2 & -1.986513 \\ \\
44 & 2 & -1.986511 \\ \\
45 & 2 & -1.986511 \\ \\
46 & 2 & -0.783959 \\ \\
47 & 2 & -0.605173 \\ \\
48 & 2 & -0.373232 \\ \\
49 & 2 & -0.356724 \\ \\
50 & 2 & -0.356724 \\ \\
51 & 2 & -0.272252 \\ \\
52 & 2 & -0.272252 \\ \\
53 & 0 & 0.132448 \\ \\
 \bottomrule
\end{tabular}
\end{table*}

\FloatBarrier
\subsection{Molecular Orbitals}
Molecular orbitals used for the symmetry analysis of each single and double excitation in the first reaction.

\begin{table*}[h]
\caption{\label{tab:table4}MO of \ce{F2}}
\begin{tabular}{cccccccc}\toprule
\\
 \multirow{1}{1cm}{Atom} & \multirow{1}{1.5cm}{Orbital} & \multirow{1}{2cm}{MO-0} & \multirow{1}{2cm}{MO-1} & \multirow{1}{2cm}{MO-2} & \multirow{1}{2cm}{MO-3} & \multirow{1}{2cm}{MO-4} & \multirow{1}{2cm}{MO-05} \\
 \\
 \hline
 \\
  0F&   1s  &       0.703304&  0.703833& -0.174874&  0.191306& -0.000000&  0.000000 \\ \\
  0F&   2s  &       0.016067&  0.012853&  0.653065& -0.766255&  0.000000& -0.000000 \\ \\
  0F&   1pz &       0.000000&  0.000000& -0.000000&  0.000000&  0.684225&  0.003020 \\ \\
  0F&   1px &       0.002716& -0.000488&  0.102207&  0.082094&  0.000140& -0.031610 \\ \\
  0F&   1py &       0.000126& -0.000023&  0.004727&  0.003797& -0.003017&  0.683495 \\ \\ 
  1F&   1s  &      -0.703304&  0.703833& -0.174874& -0.191306&  0.000000& -0.000000 \\ \\
  1F&   2s  &      -0.016067&  0.012853&  0.653065&  0.766255& -0.000000&  0.000000 \\ \\
  1F&   1pz &       0.000000& -0.000000&  0.000000&  0.000000&  0.684225&  0.003020 \\ \\
  1F&   1px &       0.002716&  0.000488& -0.102207&  0.082094&  0.000140& -0.031610 \\ \\
  1F&   1py &       0.000126&  0.000023& -0.004727&  0.003797& -0.003017&  0.683495 \\ \\
\midrule
\\
Atom& Orbital& MO-6& MO-7& MO-8 &MO-9&& \\ \\
\midrule
\\
  0F&   1s  &      -0.046464&  0.000000&  0.000000&  0.053540&& \\ \\
  0F&   2s  &       0.216449& -0.000000& -0.000001& -0.267684&& \\ \\
  0F&   1pz &       0.000000& -0.732441&  0.000053& -0.000000&& \\ \\
  0F&   1px &      -0.640713& -0.000002& -0.033835& -0.817010&& \\ \\
  0F&   1py &      -0.029629&  0.000053&  0.731659& -0.037785&& \\ \\
  1F&   1s  &      -0.046464&  0.000000&  0.000000& -0.053540&& \\ \\
  1F&   2s  &       0.216449& -0.000000& -0.000001&  0.267684&& \\ \\
  1F&   1pz &      -0.000000&  0.732441& -0.000053& -0.000000&& \\ \\
  1F&   1px &       0.640713&  0.000002&  0.033835& -0.817010&& \\ \\
  1F&   1py &       0.029629& -0.000053& -0.731659& -0.037785&& \\ \\
 \bottomrule
\end{tabular}
\end{table*}

\begin{table*}[h]
\caption{\label{tab:table4}MO of $HF$}
\begin{tabular}{cccccccc}\toprule
\\
 \multirow{1}{1cm}{Atom} & \multirow{1}{1.5cm}{Orbital} & \multirow{1}{2cm}{MO-0} & \multirow{1}{2cm}{MO-1} & \multirow{1}{2cm}{MO-2} & \multirow{1}{2cm}{MO-3} & \multirow{1}{2cm}{MO-4} & \multirow{1}{2cm}{MO-05} \\
 \\
 \hline
 \\
  0F&   1s&        -0.994789&  0.252076&  0.074062& -0.000000&  0.000000&  0.076338 \\ \\
  0F &  2s &       -0.021925& -0.956314& -0.387289&  0.000000& -0.000000& -0.475852 \\ \\
  0F  & 1pz &      -0.002412& -0.067753&  0.694635& -0.000000&  0.000000& -0.808593 \\ \\
  0F&   1px  &      0.000000& -0.000000&  0.000000&  0.214851& -0.976647& -0.000000 \\ \\
  0F &  1py   &    -0.000000&  0.000000& -0.000000& -0.976647& -0.214851& -0.000000 \\ \\
  1H  & 1s     &    0.004934& -0.142353&  0.545015& -0.000000&  0.000000&  1.021866 \\ \\
 \bottomrule
\end{tabular}
\end{table*}

\FloatBarrier
\clearpage

